\documentclass[
    aps,
    prx,
    letterpaper,
    nobalancelastpage,
    twocolumn,
    superscriptaddress,
    nofootinbib,
    longbibliography
]{revtex4-2}

\usepackage{graphicx}
\usepackage{amsmath}
\usepackage{amssymb}
\usepackage[english]{babel}
\usepackage{color}
\usepackage[version=4]{mhchem}
\usepackage[hidelinks]{hyperref}
\usepackage{dsfont}
\usepackage{multirow}
\usepackage{booktabs}
\usepackage{array}

\usepackage{soul}
\newcommand{\abs}[1]{\left| #1 \right|}

\newcommand{\ket}[1]{{| {#1} \rangle}}
\newcommand{\bra}[1]{{\langle {#1} |}}

\newcommand{\ketbra}[2]{| {#1} \rangle \langle {#2} |}

\newcommand{\dd}[1]{\,\t{d}{#1}\,}

\renewcommand{\t}[1]{\text{#1}}

\newcommand{\UIUC}{
    Department of Physics,
    The University of Illinois at Urbana-Champaign,
    Urbana, IL 61801, USA
}

\newcommand{\spamstar}[1]{\ensuremath{{}^*\mkern-2.5mu{#1}}}

\newcommand{\updated}[1]{\textcolor{black}{#1}}
\renewcommand{\cite}[1]{\mbox{\citep{#1}}}

\addto\captionsenglish{}
\addto\captionsenglish{}

\newcommand{\fillarray}{\updated{0.67(4)}}
\newcommand{\fillarraycorr}{\updated{0.68(4)}}

\newcommand{\fidelitysingle}{\updated{0.992(4)}}

\newcommand{\fidelityarray}{\updated{0.993(4)}}
\newcommand{\fidelityDarray}{\updated{0.997(2)}}
\newcommand{\fidelityBarray}{\updated{0.991(5)}}

\newcommand{\tausurvvac}{8.8(3)\,\t{s}}
\newcommand{\tausurvBarray}{1.24(8)\,\t{s}}

\newcommand{\tausurvDarray}{2.99(6)\,\t{s}}

\newcommand{\survBsingle}{\updated{0.966(8)}}
\newcommand{\survBsinglecorr}{\updated{0.99(1)}}
\newcommand{\survDsingle}{\updated{0.9963(2)}}
\newcommand{\survsingle}{\updated{0.981(4)}}
\newcommand{\survsinglecorr}{\updated{0.993(5)}}
\newcommand{\survBarray}{\updated{0.960(9)}}
\newcommand{\survBarraycorr}{\updated{0.99(1)}}
\newcommand{\survDarray}{\updated{0.9960(1)}}
\newcommand{\survarray}{\updated{0.978(5)}}
\newcommand{\survarraycorr}{\updated{0.994(3)}}

\newcommand{\survBarraycorrshift}{\updated{0.01(1)}}
\newcommand{\survDarraycorrshift}{\updated{0.0040(1)}}

\newcommand{\taudepolarray}{\updated{0.65(4)\,\t{s}}}

\newcommand{\depolDBsingle}{\updated{0.011(3)}}
\newcommand{\depolDBsinglecorr}{\updated{0.003(3)}}
\newcommand{\depolBDsingle}{\updated{0.018(4)}}
\newcommand{\depolBDsinglecorr}{\updated{0.005(7)}}
\newcommand{\depolsingle}{\updated{0.014(2)}}
\newcommand{\depolsinglecorr}{\updated{0.004(4)}}
\newcommand{\depolDBarray}{\updated{0.025(2)}}
\newcommand{\depolDBarraycorr}{\updated{0.010(6)}}
\newcommand{\depolBDarray}{\updated{0.025(2)}}
\newcommand{\depolBDarraycorr}{\updated{0.013(2)}}
\newcommand{\depolarray}{\updated{0.025(2)}}
\newcommand{\depolarraycorr}{\updated{0.012(3)}}

\newcommand{\pumpeffsingle}{\updated{0.972(9)}}
\newcommand{\pumpeffsinglecorr}{\updated{0.98(1)}}
\newcommand{\pumpeffarray}{\updated{0.972(9)}}

\newcommand{\pifidelity}{\updated{0.984(8)}}
\newcommand{\pifidelitycorr}{\updated{0.997(4)}}

\newcommand{\taudephase}{0.37(1)\,\t{s}}
\newcommand{\taudephaseecho}{1.40(5)\,\t{s}}

\begin{document}

\title{Repetitive readout and real-time control of nuclear spin qubits in $^{171}$Yb atoms}
\author{William Huie}
\affiliation{\UIUC}
\author{Lintao Li}
\affiliation{\UIUC}
\author{Neville Chen}
\affiliation{\UIUC}
\author{Xiye Hu}
\affiliation{\UIUC}
\author{Zhubing Jia}
\affiliation{\UIUC}
\author{Won Kyu Calvin Sun}
\affiliation{\UIUC}
\author{Jacob P. Covey}\email{jcovey@illinois.edu}
\affiliation{\UIUC}

\begin{abstract}
We demonstrate high fidelity repetitive measurements of nuclear spin qubits in an array of neutral ytterbium-171 ($^{171}$Yb) atoms. We show that the qubit state can be measured with a spin-flip probability of $\depolsinglecorr$ for a single tweezer and $\depolarraycorr$ averaged over the array. This is accomplished by near-perfect cyclicity of one of the nuclear spin qubit states with an optically excited state under a magnetic field of $B=58$ G, resulting in a spin-flip probability of $\approx10^{-5}$ per photon during fluorescence readout. The performance improves further as $\sim1/B^2$. The state discrimination fidelity is $\fidelityarray$ with a state-averaged readout survival of $\survarraycorr$, limited by off-resonant scattering to dark states. We combine our measurement technique with high-fidelity rotations of the nuclear spin qubit via an AC magnetic field to explore two paradigmatic scenarios, including the non-commutivity of measurements in orthogonal bases, and the quantum Zeno mechanism in which measurements ``freeze'' coherent evolution. Finally, we employ real-time feedforward to repetitively and deterministically prepare the qubit in the $+z$ or $-z$ direction after initializing it in an orthogonal basis and performing a measurement in the $Z$-basis. These capabilities constitute an important step towards adaptive quantum circuits with atom arrays.
\end{abstract}
\maketitle

\section{Introduction}\label{Intro}
Measurements play a crucial role in quantum information science to determine the result of the intended operation, correct errors~\cite{Knill1998,Knill2005}, and prepare useful many-body states~\cite{Bennett1993,Olmschenk2009, Verresen2021,Tantivasadakarn2021, Schuster2022,Lee2022,Iqbal2023,FossFeig2023}. Ideally, the qubits are not lost as a result of these measurements and remain in the state corresponding to the measured outcome. These conditions taken together constitute a quantum nondemolition (QND) measurement, which has been demonstrated on many quantum hardware platforms~\cite{Caves1980,Imoto1985,Peil1999,Hume2007, Guerlin2007,Lupascu2007,Sarovar2008,Neumann2010,Raha2019, Nakajima2019,Yoneda2020, Stricker2020,Xue2020, Yang2020,Zhang2021}. The ability to perform measurements in isolated atoms or atom-like systems in a solid state host is often hampered by the complex, multi-level structure of the atomic system. Such measurements are typically performed via optical fluorescence readout in which one qubit state is ``bright'' while the other is ``dark''. In many cases, imperfect cyclicity of the bright state leads to leakage between the two states and limits the brightness or contrast of the qubit, which can be addressed by either using single photon detectors~\cite{Robledo2011,Fuhrmanek2011,Gibbons2011,Shea2020,Chow2022} or by coupling the qubit to an optical cavity~\cite{Barclay2009,Albrecht2013,Gehr2010,Bochmann2010,Didos2018,Deist2022b}. However, neither approach is readily compatible with scalable parallel qubit readout.

\begin{figure}[t!]
    \centering
    \includegraphics[width=0.48\textwidth]{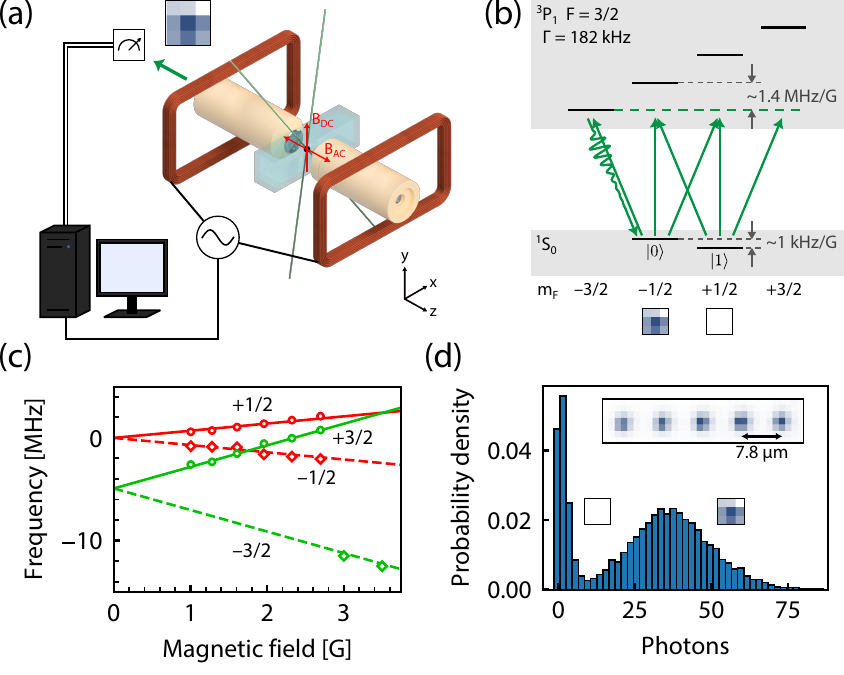}
    \caption{
        \textbf{Overview.} (a) The experimental system consists of a glass vacuum cell with one microscope objective on either side. Atoms are illuminated with two retro-reflected probe beams that each have an angle with respect to the $xy$- and $yz$-planes. The tweezer array lies parallel to the $y$-axis. The DC magnetic field and the tweezer electric field point in the $y$-direction. An AC magnetic field produced by the pair of coils shown points in the $z$-direction. Images recorded on the camera can be analyzed in real-time to enable or disable AC pulses on the coils. (b) The relevant level structure of ${}^{171}\t{Yb}$, showing the cycling transition from the ${}^1\t{S}_0$ $\ket{m_F = -1/2} \equiv \ket{0}$ qubit state making it bright while $\ket{m_F = +1/2} \equiv \ket{1}$ is dark during readout. (c) Energies of the four $m_F$ Zeeman states in ${}^3\t{P}_1$ $F = 3/2$ relative to free space versus magnetic field, taken with a $\approx 1\,\t{mK}$ tweezer depth for a single array site. (d) A histogram of \updated{$12\,\t{ms}$} fluorescence readout (\updated{$2500\,\t{repetitions} \times 5\,\t{sites} = 12500\,\t{shots}$}) of the nuclear spin qubit,  where the dark peak is either zero atoms or an atom in $|1\rangle$ and the bright peak is an atom in $|0\rangle$. The system was initialized in $|0\rangle$. The discrimination fidelity is $\mathcal{F} = \fidelityarray$. Inset: Averaged (\updated{$2500\,\t{repetitions}$}) camera image of fluorescence from a 5-site array with spacing of $7.8\,\t{$\mu$m}$.
        \label{Figure1}
    }
\end{figure}

Arrays of neutral atoms in optical tweezers~\cite{Kaufman2021} are rapidly emerging as a leading platform for myriad quantum science applications ranging from quantum simulation~\cite{Browaeys2020,Morgado2021}, computing~\cite{Saffman2010,Morgado2021}, and sensing~\cite{Norcia2019,Madjarov2019} to networking~\cite{Reiserer2015,Covey2023}. Scalable, lossless readout of hyperfine qubits has been performed with arrays of neutral alkali atoms in optical tweezers~\cite{MartinezDorantes2017,Kwon2017,Nikolov2023}, and QND readout of hyperfine~\cite{Kuzmich2000,SchleierSmith2010} and nuclear spin qubits~\cite{Braverman2019,Zheng2022,Yang2023} has been performed with ensembles of atoms in an optical dipole trap. The use of optical qubits helps obviate cyclicity limitations and thus optical qubits are readily compatible with scalable single-atom QND measurements~\cite{Liebfried2003,Hume2007,Covey2019,Norcia2019,Madjarov2019}, but are hampered by requirements on optical phase stability and atomic temperature. For spin-encoded qubits on the other hand, a scalable approach to QND readout in arrays of neutral atoms in optical tweezers remains an outstanding challenge. Namely, $\approx1000$ photons must be scattered for high fidelity detection of single atoms in free space with collection efficiencies typically at the single-percent level, and thus QND readout with percent-level atom loss and qubit depolarization requires such events to be at the $\approx10^{-5}$ level per photon.

Here, we leverage the unique atomic structure of neutral $^{171}$Yb atoms to directly perform high-fidelity QND measurements of a qubit encoded in the nuclear spin-1/2 degree of freedom in the electronic ground state. By performing fluorescence detection via the relatively narrow $^3$P$_1$ optically excited state in which the $m_F=-3/2$ Zeeman sub-level is sufficiently isolated at a modest magnetic field of $58$ G, the polarization selection rule for decay to only the $m_F=-1/2$ ground state provides a cyclicity of $\approx10^5$ with respect to the $m_F=+1/2$ ground state, corresponding to an average qubit depolarization probability of $\depolsinglecorr$ for a single tweezer ($\depolarraycorr$ for array-averaged) during $12$-ms measurements with fidelity of $\fidelityarray$ and survival probability of $\survarraycorr$. We demonstrate this technique for an array of $^{171}$Yb atoms in optical tweezers of wavelength $\approx760\,\t{nm}$ -- ideal for subsequent manipulation of the optical ``clock'' transition~\cite{Ye2008,Ludlow2015}. Unlike the seminal demonstrations of free space non-destructive qubit readout in alkali atom arrays~\cite{MartinezDorantes2017,Kwon2017,Nikolov2023}, our tweezers are relatively shallow ($U_0/k_B\approx 580\,\t{$\mu$K}$) and remain on the entire time, obviating the need to chop them out of phase with the probe light. Moreover, the probe beams are randomly polarized and have projections onto all three dimensions, allowing the atoms to stay cold in three dimensions under probe illumination, with a temperature of $T\approx5$ $\mu$K. 

We combine high-fidelity projective measurements with qubit rotations to explore textbook scenarios including observation of the non-commutivity of measurements in variable bases, and demonstration of the quantum Zeno mechanism by studying the interplay of measurement and qubit rotation during repetitive alternation. Finally, we implement real-time adaptive control~\cite{Singh2022} to perform a qubit rotation conditioned on the measurement outcome in order to deterministically prepare a target state after a projective measurement, and we show the ability to repetitively do so in alternation with a rotation to an orthogonal basis. If combined with the ability to perform measurements on only subsets of qubits~\cite{Deist2022b,Singh2022,Graham2023}, this work would aid in the realization of measurement-based quantum computation~\cite{Raussendorf2001,Walther2005,Stephen2022}, non-unitary many-body state preparation protocols~\cite{Verresen2021,Tantivasadakarn2021,Lee2022,Iqbal2023,FossFeig2023}, quantum error correction~\cite{Knill1998,Knill2005}, and the study of measurement-induced phase transitions~\cite{Skinner2019,Bao2019,Gullans2020b,Noel2022,Mingkoh2022}.       

\section{Overview of the experimental system}\label{system}
We begin with a laser cooled ensemble of $^{171}$Yb atoms suspended in the center of a glass cell held under ultrahigh vacuum [see Fig.~\ref{Figure1}(a)]. High-resolution microscope objectives with diffraction-limited $\t{NA}\approx0.6$ are placed on either side of the glass cell. A one-dimensional array of five optical tweezers with wavelength $\lambda_T\approx760\,\t{nm}$ and $1/e^2$ waist radius of $w_0\approx670$ nm is generated with an acousto-optic deflector (AOD)~\cite{Endres2016}. The spacing between adjacent tweezers is $d = 7.8\,\t{$\mu$m}$, corresponding to a frequency difference between adjacent radio frequency tones sent to the AOD of $1.75$ MHz. The inset to Fig.~\ref{Figure1}(d) shows $2500$-shot averaged images of the fluorescence from our $5$-site array. We use a power of $7\,\t{mW}$ per tweezer, which corresponds to a depth of $U_0/h\approx12\,\t{MHz}$ ($U_0/k_B\approx580\,\t{$\mu$K}$) in the ground state. Appendix~\ref{apparatus} provides further details of our experimental system.

\begin{figure}[b!]
    \centering
    \includegraphics[width=0.48\textwidth]{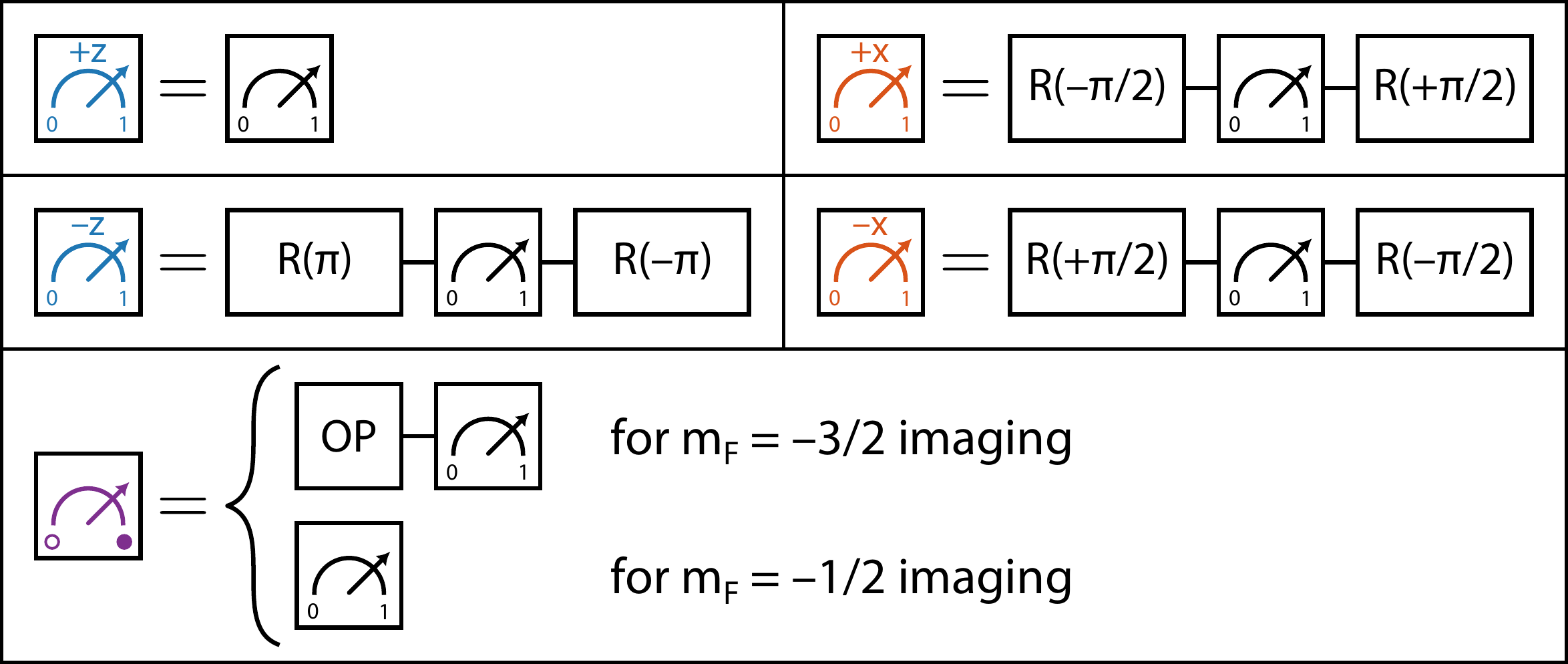}
    \caption{
        \textbf{Circuit legend.} Simplified equivalent circuit elements. The number 0/1 and the open/filled circle distinguishes ``qubit readout'' from ``atom readout'', respectively. 
        \label{Figure2}
    }
\end{figure}

\begin{figure*}[t!]
	\centering
	\includegraphics[width=0.8\textwidth]{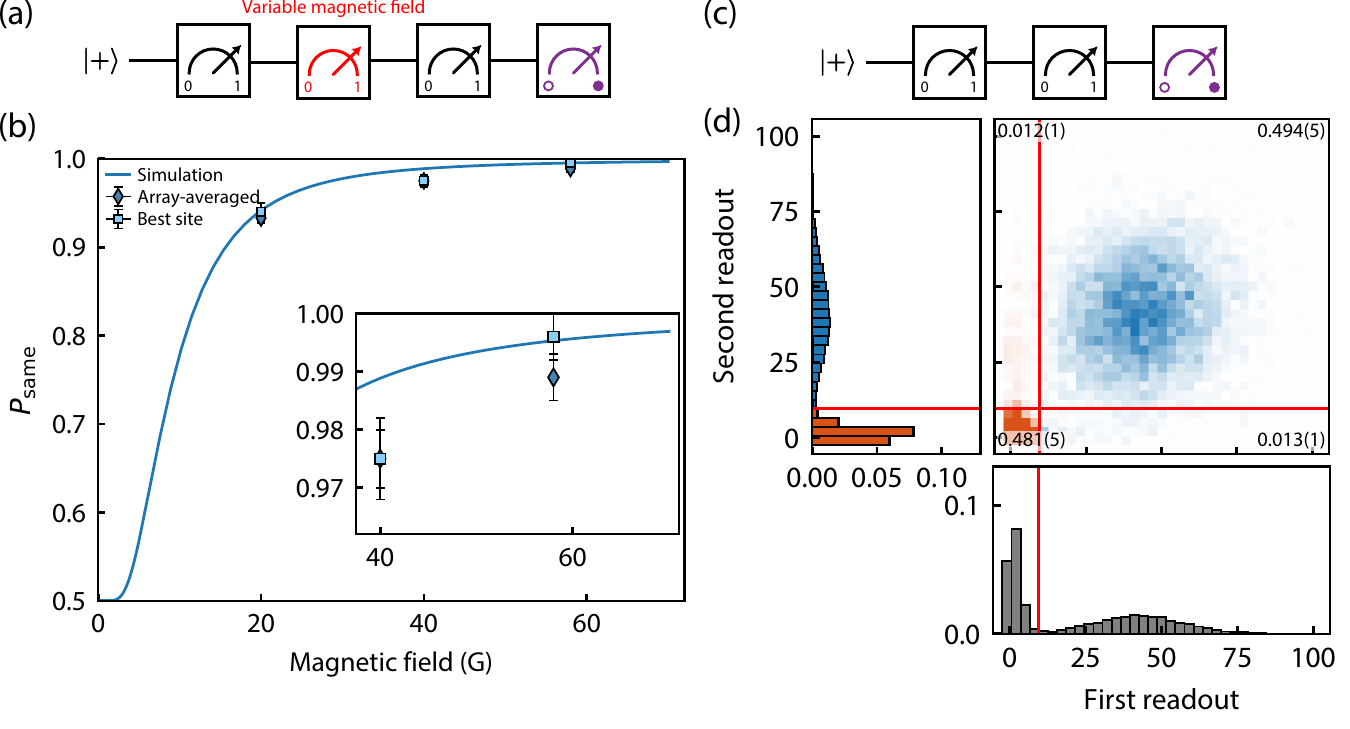}
    \caption{
        \textbf{Characterizing depolarization during qubit readout}. (a) Circuit diagram for studying the dependence of depolarization during qubit readout on magnetic field. A probe block of variable magnetic field is placed between two measurements performed at $58\,\t{G}$. The final ``atom readout'' pulse is used to post-select on events in which the atom survived the entire sequence. (b) The measured array-average (diamonds) and single-site best (squares), SPAM-corrected, and the predicted state populations (line) under real imaging conditions versus magnetic field. The error bars reflect the standard deviation of a binomial distribution. (c) Circuit diagram for directly studying the depolarization probability during qubit readout at $58\,\t{G}$. (d) The array-averaged histograms from each image for \updated{$17500$} shots, with the results shown together on a 2D histogram. These results show that the D\,$\rightarrow$\,B and B\,$\rightarrow$\,D conditional depolarization probabilities during a \updated{$12\,\t{ms}$} probe pulse are $\depolDBarray$ ($\depolDBsingle$) and $\depolBDarray$ ($\depolBDsingle$) averaged across the array (best site), respectively, before SPAM correction.
        \label{Figure3}
    }
\end{figure*}

The tweezers are continuously on during the MOT phase; atoms remain in the tweezer traps after the MOT light and magnetic field gradient have been turned off. Single atoms are obtained in the tweezers by applying a cooling pulse using the MOT beams under a field of $1.5\,\t{G}$ in the $y$-direction, which is also the direction of the tweezer's polarization [see Fig.~\ref{Figure1}(a)]. This pulse has a total intensity of $I_\t{cool}=1.3\,I_\t{sat}$, where $I_\t{sat}$ is the saturation intensity of the intercombination transition, with a detuning of $\delta/2\pi\approx-150$ kHz ($\approx-0.8\,\Gamma$) from the $F=3/2$, $m_F=-1/2$ state. This cooling pulse drives light-assisted collisions that transform the initially Poissonian atom number distribution into either just 0 or 1 atom remaining~\cite{Schlosser2001}. After the cooling pulse, we obtain a tweezer loading fraction of \updated{$p\approx0.7$} -- which is correlated with the density of the reservoir from which we load, suggesting that higher loading fractions are possible~\cite{Jenkins2022} -- and an atomic temperature of $T\approx5\,\t{$\mu$K}$ measured via release and recapture from the tweezers~\cite{Tuchendler2008}. See Appendices~\ref{apparatus} and~\ref{32vs12} for further details.

Readout is performed using the same transition as that for the MOT ($^1$S$_0\leftrightarrow$~$^3$P$_1$, $F=3/2$)~\cite{Saskin2019,Ma2022}, where the transition linewidth of $\Gamma/2\pi=182$ kHz is well suited for our photon scattering rate of \updated{$\Gamma_\t{scatt}\approx97000\,\t{photons/s}$} during fluorescence detection. We use two counter-propagating beams that each have an angle of $\approx15$ degrees with respect to $xy$-plane and an angle of $\approx\pm30$ degrees with respect to the $yz$-plane [see Fig.~\ref{Figure1}(a)]. This configuration is used to minimize the effect of surface scatter from the beams, which do not pass through the faces of the cell used by the microscope objectives. The polarization of the beams is chosen to have large projections onto both $\pi$ and $\sigma^\pm$. We use an electron-multiplying CCD (EMCCD) to image atomic fluorescence via the microscope objective opposite the one used to generate the tweezers. Typical readout pulses are \updated{$12\,\t{ms}$} long with probe intensity of \updated{$I_\t{probe}\approx1\,I_\t{sat}$} (the total beam intensity is $\approx3$ times higher but we assume it is equally divided among the three polarizations) and detuning \updated{$\delta/2\pi\approx-180\,\t{kHz}$} with respect to the target $m_F$ state within the $F=3/2$ manifold. Our imaging system uses a magnification of $\approx9$, and our estimated atom-to-camera collection efficiency is \updated{$\approx0.04$} based on the calculated scattering rate for these probe conditions and the number of photons collected on the camera.

\begin{table*}[t!]
    \renewcommand*{\arraystretch}{1.5}
    \centering
    \caption{
        \textbf{Summary of discrimination fidelity, atom survival, and
        depolarization probabilities}.
        The discrimination fidelity $\mathcal{F}$, atom survival probabilities
        $\eta_\t{surv}^\t{B}$, $\eta_\t{surv}^\t{D}$, $\bar{\eta}_\t{surv}$, and
        depolarization probabilities $P_\t{depol}^{\t{D}\,\rightarrow\,\t{B}}$,
        $P_\t{depol}^{\t{B}\,\rightarrow\,\t{D}}$, $\bar{P}_\t{depol}$ are
        listed for the $m_F = -3/2$ imaging condition both on a single array
        site and averaged over a 5-site array. Symbols with superscripts are
        state-dependent, with `B' referring to the $\ket{0}$ bright state and
        `D' referring to the $\ket{1}$ dark state, while those with bars are
        state-averaged. Numbers are listed with and without SPAM
        correction.
        \label{numbers}
    }
    \begin{tabular}{
        l<{\quad}
        l<{\quad}
        l<{\quad}
        l<{\quad}
        l<{\quad}
        l<{\quad}
        l<{\quad}
        l
    }
        \hline\hline
        \multicolumn{8}{c}{Array-averaged} \\
        \hline
        \quad
            & $\mathcal{F}$
            & $\eta_\t{surv}^\t{B}$
            & $\eta_\t{surv}^\t{D}$
            & $\bar{\eta}_\t{surv}$
            & $P_\t{depol}^{\t{D}\,\rightarrow\,\t{B}}$
            & $P_\t{depol}^{\t{B}\,\rightarrow\,\t{D}}$
            & $\bar{P}_\t{depol}$
        \\
        \hline
        Uncorrected
            & $\fidelityarray$
            & $\survBarray$
            & $\survDarray$
            & $\survarray$
            & $\depolDBarray$
            & $\depolBDarray$
            & $\depolarray$
        \\
        Corrected
            & ---
            & $\survBarraycorr$
            & ---
            & $\survarraycorr$
            & $\depolDBarraycorr$
            & $\depolBDarraycorr$
            & $\depolarraycorr$
        \\
        \hline\hline
        \multicolumn{8}{c}{Best site} \\
        \hline
        \quad
            & $\mathcal{F}$
            & $\eta_\t{surv}^\t{B}$
            & $\eta_\t{surv}^\t{D}$
            & $\bar{\eta}_\t{surv}$
            & $P_\t{depol}^{\t{D}\,\rightarrow\,\t{B}}$
            & $P_\t{depol}^{\t{B}\,\rightarrow\,\t{D}}$
            & $\bar{P}_\t{depol}$
        \\
        \hline
        Uncorrected
            & $\fidelitysingle$
            & $\survBsingle$
            & $\survDsingle$
            & $\survsingle$
            & $\depolDBsingle$
            & $\depolBDsingle$
            & $\depolsingle$
        \\
        Corrected
            & ---
            & $\survBsinglecorr$
            & ---
            & $\survsinglecorr$
            & $\depolDBsinglecorr$
            & $\depolBDsinglecorr$
            & $\depolsinglecorr$
        \\
        \hline\hline
    \end{tabular}
\end{table*}

\section{Nondestructive qubit readout}\label{readout}
A crucial feature of our nondestructive qubit readout technique is the electric dipole polarization selection rule associated with our choice of excited state [see Fig.~\ref{Figure1}(b)]. The $m_F$ states within $^3$P$_1$ are well resolved even at low magnetic fields due to the relatively narrow linewidth ($\Gamma/2\pi\approx182$ kHz) and the large g-factor ($\approx1.4$ MHz/G), as shown in Fig.~\ref{Figure1}(c). Based on the zero-field detunings of $|m_F|=1/2$ and $|m_F|=3/2$ and the known ground-state light shift, we estimate the differential polarizabilities $\alpha=(U_e-U_0)/U_0$ at this tweezer wavelength ({$\lambda_T\approx760\,\t{nm}$) to be $\alpha_{|1/2|}\approx-0.030(3)$ and $\alpha_{|3/2|}\approx0.25(3)$ (see Appendix~\ref{polarizability}), which are in good agreement with recent observations~\cite{Jenkins2022,Ma2022}. Although the $|m_F|=1/2$ states are appealing due to the nearly-zero differential light shift and the assurance that both nuclear spin ground states will remain bright~\cite{Ma2022}, the positive differential light shift of the $|m_F|=3/2$ state corresponds to the case where the excited state is deeper trapped than the ground state -- a scenario in which \textit{attractive} Sisyphus cooling has been observed for strontium (Sr)~\cite{Covey2019,Barnes2021,Urech2022} and predicted for Sr and Yb~\cite{Taieb1994,Ivanov2011}.

In this work, we focus on the $m_F=-3/2$ excited state which, under ideal conditions, can decay only to the $m_F=-1/2$ ground state. This allows us to perform ``qubit readout'' since the $|m_F=-1/2\rangle\equiv|0\rangle$ state will remain bright while the $|m_F=+1/2\rangle\equiv|1\rangle$ state is dark [see Fig.~\ref{Figure1}(b)]. It is also crucial to be able to perform ``atom readout'' -- which is state-independent -- in order to differentiate a perceived outcome of $|1\rangle$ in a qubit measurement from cases where the atom may have been lost. We employ two techniques for performing atom readout (see Fig.~\ref{Figure2}). One is to use the $m_F=-1/2$ excited state which is connected to both ground states; the other is to re-initialize the qubit in $|0\rangle$ via optical pumping and then perform qubit readout. Our optical pumping efficiency is $\pumpeffsinglecorr$ with state preparation and measurement correction (see Appendices~\ref{apparatus} \& \ref{spam}), $\pumpeffsingle$ without. We focus primarily on the latter technique, mostly to avoid the need to change the probe frequency by many tens of MHz when going between the $m_F=-3/2$ and $m_F=-1/2$ excited states. Nevertheless, we find that imaging with $m_F=-3/2$ and $-1/2$ offer similar performance: the collection efficiency is similar, and the steady-state temperature under probing is $T\approx5$ $\mu$K for both. See Appendix~\ref{32vs12} for further details on their comparison.

We note that both cases are limited by tweezer-induced off-resonant scatter during probing and cooling from the steady state population in $6s6p$ $^3$P$_1$ to the higher $6s7s$ $^3$S$_1$ state, which can then subsequently decay to the entire $6s6p$ $^3$P$_J$ manifold. In principle, $^3$P$_2$ and $^3$P$_0$ could be repumped, but we note that $^3$P$_2$ -- which is the dominant decay path -- is unfortunately strongly anti-trapped in tweezers with wavelength $\lambda_T\approx760\,\t{nm}$ due to its proximity to the $^3$P$_2\leftrightarrow$~$^3$S$_1$ transition at $770.2\,\t{nm}$. Without any repumping, we observe the lifetime in $\ket{0}$ under probing to be \updated{$\tau=\tausurvBarray$} which is consistent with our model (see Apppendix~\ref{atomloss}). We choose a probe time of \updated{$12\,\t{ms}$} as an optimal compromise that offers a bright/dark discrimination fidelity of $\mathcal{F}=\fidelityarray$ corresponding to the histogram in Fig.~\ref{Figure1}(d), and a probe survival of the $|0\rangle$ state of $\survBarraycorr$ with state preparation and measurement (SPAM) correction (see Appendix~\ref{spam}), in good agreement with our measured lifetime. Significant gains are possible by operating in a shallower tweezer and improving the collection and detection efficiency. We measure the lifetime in $|1\rangle$ under probe conditions to be \updated{$\tausurvDarray$}, suggesting that the survival of the $|1\rangle$ dark state is \updated{$\survDarray$}. The state-averaged survival during qubit readout is thus taken to be \updated{$\survarraycorr$}.

To ensure that the $|0\rangle$ qubit state remains bright while the $|1\rangle$ qubit state remains dark during probing, we require excellent isolation of the $|0\rangle\leftrightarrow|3/2,-3/2\rangle$ transition. There are two effects that limit this isolation: Raman transitions via other excited states, and mixing between the excited states. We analyze these effects in Appendices~\ref{multilevel} and~\ref{mixing}, respectively, showing that both are suppressed quadratically in magnetic field. The Raman transitions have a spontaneous contribution and a stimulated contribution. The latter is due to the presence of all polarization components [see Fig.~\ref{Figure1}(b)] but the effect is suppressed by the $\approx45$ kHz nuclear spin splitting at 58 G. Although choosing our probe polarizations to contain only $\sigma^\pm$ components would have broken the stimulated Raman condition and removed the $m_F = -1/2\rightarrow m_F' = -1/2$ channel, the results would not have significantly changed due to the inevitable $m_F = +1/2\rightarrow m_F' = -1/2$ channel [see Fig.~\ref{Figure1}(b)], and doing so would have cost us the ability to use ``atom readout'' via the $m_F = -1/2$ excited state. The mixing between the excited Zeeman states is zero with a perfectly linearly polarized tweezer whose polarization is perfectly aligned with the magnetic field. Finite mixing emerges due to deviations from this perfect case, but they are suppressed as $\sim1/B^2$ as shown in Appendix~\ref{mixing}.

We study the qubit depolarization of our $-3/2$ imaging protocol for several magnetic fields. We place a \updated{$12\,\t{ms}$} qubit readout block performed at a variable magnetic field between two qubit readout blocks performed at $B=58\,\t{G}$ [see Fig.~\ref{Figure3}(a)]. We place an atom readout block at the end of the sequence to post-select on events where the atom survives the entire sequence. As shown in Fig.~\ref{Figure3}(b), we see good agreement with the expected $\sim1/B^2$ scaling. The case of 58 G is studied in further detail in Fig.~\ref{Figure3}(c) and (d). By plotting all camera counts in each image for all $17500$ shots with respect to the dark/bright (D/B) threshold [see Fig.~\ref{Figure3}(d)], we can directly measure the probability of all four events: B\,$\rightarrow$\,B, D\,$\rightarrow$\,D, B\,$\rightarrow$\,D, and D\,$\rightarrow$\,B. These results -- with further analysis described in Appendix~\ref{analysis} -- indicate array-averaged and state-averaged conditional depolarization probabilities of $\bar{P}_\t{depol} = \depolarraycorr$ with SPAM correction; $\bar{P}_\t{depol} = \depolarray$ without. Our ability to fully control the tweezer polarization across the array is currently limited by a slight defocus in the tweezer array optics. This causes the tweezers to not be exactly parallel, which matters because the tweezer polarization is along the array axis. Therefore, we see substantially better depolarization values in the center of our array where the polarization is better aligned to $B_\t{DC}$ (see Appendix~\ref{mixing}): the state-averaged corrected and raw depolarizations for our best site are $\bar{P}_\t{depol} = \depolsinglecorr$ and $\bar{P}_\t{depol} = \depolsingle$, respectively. This issue can be addressed with straightforward adjustments to our optics or by rotating our array by 90 degrees, which we leave for future work. We have explicitly measured the array-averaged spin-flip time scale to be $\taudepolarray$, which is in good agreement with both site-by-site and array-averaged spin-flip probabilities during readout.

The measured results are listed in Table~\ref{numbers}, and include both raw and corrected values (see Appendix~\ref{analysis}) for array averages and the best site. As noted above, there is a \updated{$\survBarraycorrshift$} (\updated{$\survDarraycorrshift$}) probability of atom loss in $|0\rangle$ ($|1\rangle$) during readout, which would manifest mostly as an inflated B\,$\rightarrow$\,D probability (a raw histogram is shown in Appendix~\ref{analysis}). However, this issue can be addressed via post-selection as we have done, or in real time via a second measurement after a $\pi$-pulse on the qubit.

\begin{figure}[t!]
	\centering
	\includegraphics[width=0.48\textwidth]{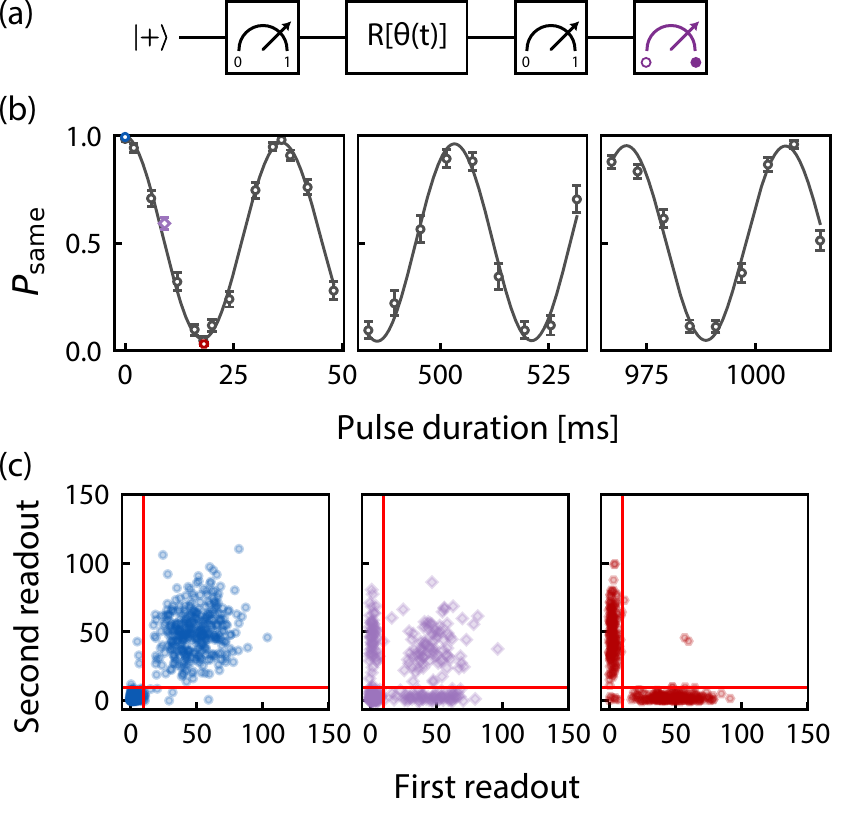}
    \caption{
        \textbf{Interleaved readout and qubit rotation}. (a) Circuit for qubit rotations between two qubit readout pulses. The atom readout pulse is included at the end to post-select on events in which the atom survives the entire sequence. We start with a state aligned along $|+\rangle=(|0\rangle+|1\rangle) / \sqrt{2}$. (b) Rabi oscillations versus time, showing three groups that span the first second. No obvious contrast decay is observed. $P_\t{same}$ refers to cases where the outcome of the two qubit measurements are the same. This outcome is equally split between $|0\rangle$ and $|1\rangle$ due to the initial state $|+\rangle$. (c) The 2D scatter plots of photon counts associated with the early-time data points with rotations of $0$, $\approx\pi/2$, and $\pi$ as shown in (b). These results indicate that the $\pi$-pulse fidelity is comparable to the probability of not undergoing a spin flip during readout. 
        \label{Figure4}
    }
\end{figure} 

\section{Interleaved readout and qubit rotation}\label{rotate}
We now add qubit rotations to demonstrate the utility of high-fidelity repetitive qubit readout. Rotations are driven by an AC magnetic field perpendicular to the DC field. At $B_\t{DC}=58$ G, the nuclear spin qubit splitting is $f \approx 43.5\,\t{kHz}$. We apply up to $B_\t{AC}=0.29$ G directly to our shim coil pair in the $x$-direction [see Fig.~\ref{Figure1}(a)], for which the Rabi frequency is $\Omega_\t{RF}/2\pi\approx100$ Hz (see Appendix~\ref{acfield} for further details). Similar results were recently obtained with a designated antenna loop~\cite{Ma2022}. The data shown below uses a Rabi frequency of $\Omega_\t{RF}/2\pi\approx28$ Hz to mitigate transient effects associated with the AC field. We note that stimulated Raman rotations via optical transitions offer Rabi frequencies on the $\sim$MHz scale~\cite{Barnes2021,Jenkins2022}. 

\textbf{Rabi and Ramsey coherence}. We add qubit rotations between two qubit readout pulses as shown in Fig.~\ref{Figure4}(a). We start with a state polarized along $|+\rangle=(|0\rangle+|1\rangle)/\sqrt{2}$ using a $\pi/2$-pulse, such that the probabilities of measuring $|0\rangle$ and $|1\rangle$ in the first image are equal. We plot the probability that both image outcomes are the same, $P_\t{same}$, as shown in Fig.~\ref{Figure4}(b). No obvious contrast decay is observed in Rabi oscillations extending out to one second. With the pulse sandwiched between two qubit readout blocks, we show a scatter plot of counts in both images for all $200$ shots [see Fig.~\ref{Figure4}(c)]. We show plots for rotations of $\theta=0$, $\approx\pi/2$, and $\pi$. In the case of the $\pi$-pulse, we see that nearly all occurrences are B\,$\rightarrow$\,D and D\,$\rightarrow$\,B, and thereby calculate the $\pi$-pulse fidelity as $\pifidelitycorr$ with correction and $\pifidelity$ without (see Appendix~\ref{spam}).

We also perform a Ramsey sequence to characterize the $T_2^*$. We use two resonant $\pi/2$ pulses separated by a dark time $\tau$ and we vary the phase of the second pulse to obtain a Ramsey fringe. We plot the fringe contrast versus $\tau$ to extract $T_2^*$. See Appendix~\ref{T1T2*}. The array-averaged contrast data is well described by a Gaussian envelope with $1/e$ contrast occurring at $T_2^* = \taudephase$. We observe similar values at 30 and 90 G (within $\approx25$\%), and we believe that we are limited by ambient magnetic field noise (see Appendix~\ref{T1T2*}). We note that $T_2^*=0.7(3)$ s has been realized with molecular nuclear spins at 86 G~\cite{Park2017} and that $\approx2$ mG stability has been realized at $\approx1000$ G~\cite{Borkowski2023}. By adding an echo pulse in the Ramsey sequence, we observe an extended coherence time of $T_2^\t{echo} = \taudephaseecho$ (see Appendix~\ref{T1T2*}). We also observe a $1/e$ $T_1$ time of $\approx200$ seconds (see Appendix~\ref{T1T2*}). This is somewhat longer than values reported at low field, but is consistent with observed trends as the field is increased~\cite{Jenkins2022} due to the reduced spectral weight of magnetic field noise near the qubit frequency -- a convenient feature of operating at higher field. We note that our non-destructive readout technique makes it straightforward to differentiate between atom loss and spin-flip events when measuring $T_1$.

\begin{figure}[t!]
	\centering
	\includegraphics[width=0.48\textwidth]{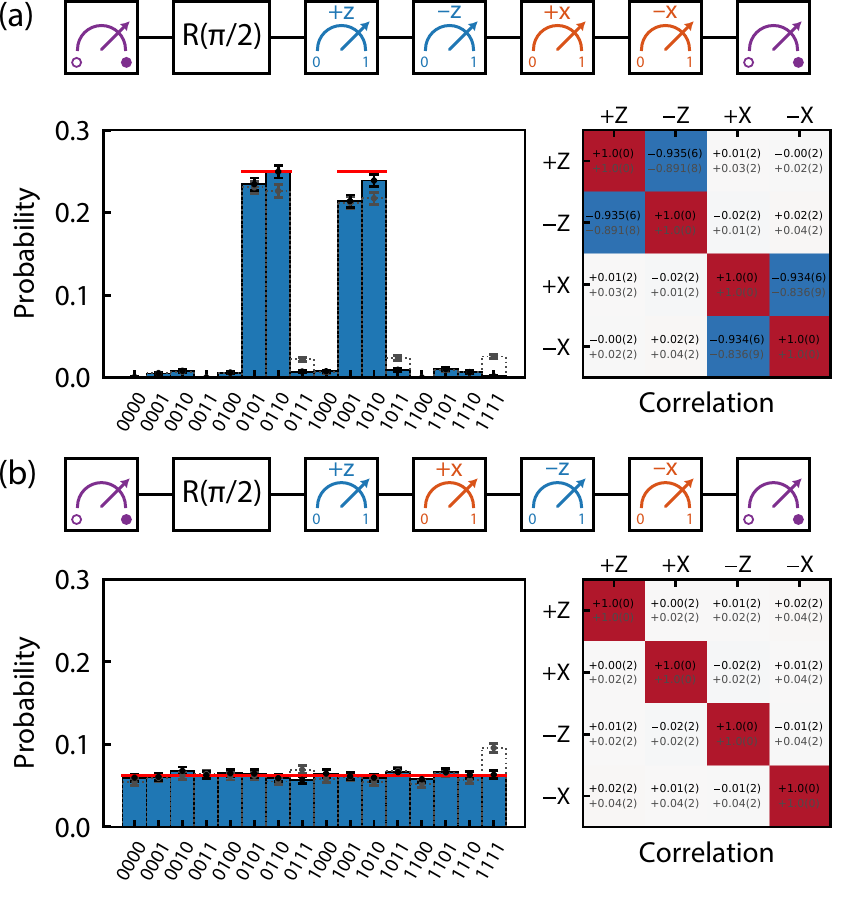}
    \caption{
        \textbf{Repetitive readout in variable bases}. (a) The $\{+Z,-Z,+X,-X\}$ measurement sequence, showing large bit string probabilities only for cases where the outcome is opposite between both the first two and last two measurements. The correlation matrix shows strong off-diagonal negative correlations for these pairs. (b) The $\{+Z,+X,-Z,-X\}$ measurement sequence, showing equal bit string probabilities for all outcomes. The correlation matrix shows no significant off-diagonal elements. The red lines show the ideal probability distributions, which are either $0$ or $1/2^2=0.25$ in (a) and $1/2^4=0.0625$ in (b). The ideal off-diagonal correlations are either $0$ or $-1$. Data in black and solid bars post-selects on detecting an atom in the initial and final atom readout; data in gray and dotted bars post-selects on only the initial atom readout.
        \label{Figure5}
    }
\end{figure}  

\textbf{Repetitive readout in variable bases}. The non-commutivity of measurements in different bases is a hallmark feature of quantum behavior and underlies the textbook examples of cascaded Stern-Gerlach devices and optical polarizers. To further show the unique capabilities of our repetitive qubit readout technique, we conduct a version of such experiments by recognizing that readout in any fixed basis can be combined with qubit rotations to perform readout in any other basis: We rotate the desired axis of the Bloch sphere into the measurement direction (we define this without loss of generality to be in the $z$-basis with a bright count mapped to $+Z = \ket{0}$) and then rotate back (see Fig.~\ref{Figure2}). We compare two measurement sequences: $\{+Z,-Z,+X,-X\}$ and $\{+Z,+X,-Z,-X\}$. Figure~\ref{Figure5}(a) and (b) show these two cases, respectively, where the qubit has again been initialized in $|+\rangle$. We show histograms of the outcome bit strings, where the anti-correlation between the first two and latter two measurements is clearly apparent in the first sequence, while the non-commutativity between each consecutive measurement in the second sequence leads to the absence of correlations. We also directly quantify these correlations through a correlation matrix that shows strong, negative off-diagonal elements in the former case, but only diagonal elements in the latter case as expected. We show the results with and without post-selection on detecting an atom at the end of the sequence; atom loss during the sequence biases measurement outcomes toward bit strings ending with $1$'s.

\textbf{The quantum Zeno mechanism}. We can also study the interplay between qubit rotations and projective measurements. The quantum Zeno mechanism describes the scenario where the measurement rate is large compared to the qubit rotation rate, such that projection back to the initial, un-rotated state overwhelms the growth of population in others. This behavior has been observed in myriad experimental systems~\cite{Streed2006,Syassen2008,Zhu2014,Patil2015}, and we use our ability to interleave qubit readout and rotation to access this regime in a unique and discrete manner. 

\begin{figure}[t!]
	\centering
	\includegraphics[width=0.48\textwidth]{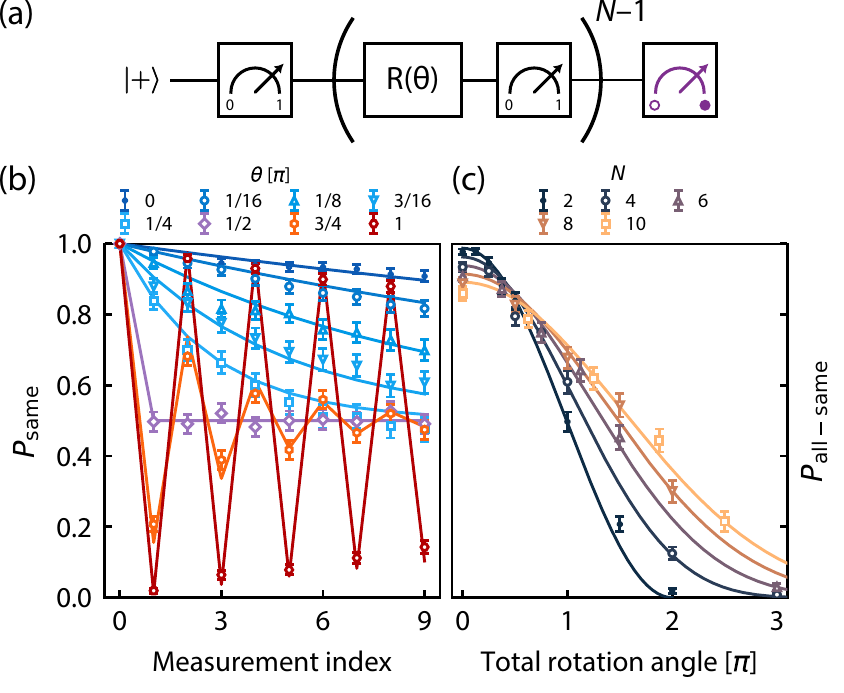}
    \caption{\textbf{The quantum Zeno mechanism}. (a) The Zeno circuit. After initializing in $|+\rangle$ and performing a qubit measurement, we repetitively interleave a rotation $R(\theta)$ and a readout $N-1$ times. We post-select on the final atom readout. (b) The probability that the measurement has the same outcome as the initial measurement, $P_\t{same}$, versus the measurement index for many different rotation angles $\theta$. We see monotonic decay of $P_\t{same}$ to 0.5 for $\theta \in [0, \pi/2]$ and damped alternations to 0.5 for $\theta \in [\pi/2, \pi]$. The lines deviate from the ideal case only by the finite depolarization probability during readout which is fit to the $\theta = 0$ case as $\bar{P}_\t{depol}=0.0127(6)$. (c) The probability that all outcomes are the same, $P_\t{all-same}$, versus total rotation angle $\theta_\t{tot}=\theta\times(N-1)$ for several different numbers of repetitions $N$. The lines are $P_\t{all-same}=(1-\bar{P}_\t{depol})^N\t{cos}^{2N-2}(\theta_\t{tot}/(2 N - 2))$.
        \label{Figure6}
    }
\end{figure} 

Specifically, after initializing the atom in $\ket{+}$ and performing a first qubit readout, we apply alternating qubit rotations $R(\theta)$ and readout for $N = 10$ times for variable rotation angle $\theta$ [see Fig.~\ref{Figure6}(a)]. We plot the average probability of finding the qubit to be in the same state as the first readout, $P_\t{same}$, versus the image index for the $N$ images. When $\theta=0$, we expect to always obtain $P_\t{same}=1$. The observed slow decay is due to the small but finite depolarization probability $\bar{P}_\t{depol}$ in each image; $P_\t{same}(N=10)\approx0.9$ is in good agreement with $(1-\bar{P}_\t{depol})^N$ for $\bar{P}_\t{depol}\approx0.01$. When $\theta=\pi$, the measurement outcome should alternate between $|0\rangle$ and $|1\rangle$ with contrast limited only by $\bar{P}_\t{depol}$ and the $\pi$-pulse fidelity, consistent with our observations. For intermediate angles, the probabilistic nature of outcomes in each measurement combined with the averaging over all trajectories leads to a damping of $P_\t{same}$ that asymptotically approaches $P_\t{same}=0.5$. Figure~\ref{Figure6}(b) shows these expected trends, where values within $\theta \in [0, \pi/2]$ decay monotonically to $P_\t{same}=0.5$ and values within $\theta \in [\pi/2, \pi]$ undergo damped alternation.

The quantum Zeno mechanism illustrates how projective measurements can suppress qubit dynamics when the measurement rate exceeds the coherent qubit rotation rate. Accordingly, for a total rotation angle $\theta_\t{tot}$ applied to the qubit, the Zeno mechanism predicts a strong dependence of the dynamics on the number $N-1$ of projective measurements during the rotation. Specifically, the rotation angle $\theta_\t{tot}/(N-1)$ between each measurement leaves the qubit in its initial state with probability $P_\t{same}=\t{cos}^2(\theta_\t{tot}/(2 N - 2))$. The probability that the qubit has remained in the state measured in the first readout during all $N-1$ subsequent measurements is therefore $P_\t{all-same}=P_\t{same}^{N-1}= \t{cos}^{2N-2}(\theta_\t{tot}/(2 N - 2))$. Figure~\ref{Figure6}(c) shows this trend, limited primarily by the (uncorrected) depolarization probability $\bar{P}_\t{depol}\approx0.01$ for each measurement. The data is in good agreement with $(1-\bar{P}_\t{depol})^N\times P_\t{all-same}$, and captures the essence of the Zeno mechanism in which $P_\t{all-same}\rightarrow1$ as $N\rightarrow\infty$ for fixed $\theta_\t{tot}$.

\begin{figure}[t!]
	\centering
	\includegraphics[width=0.48\textwidth]{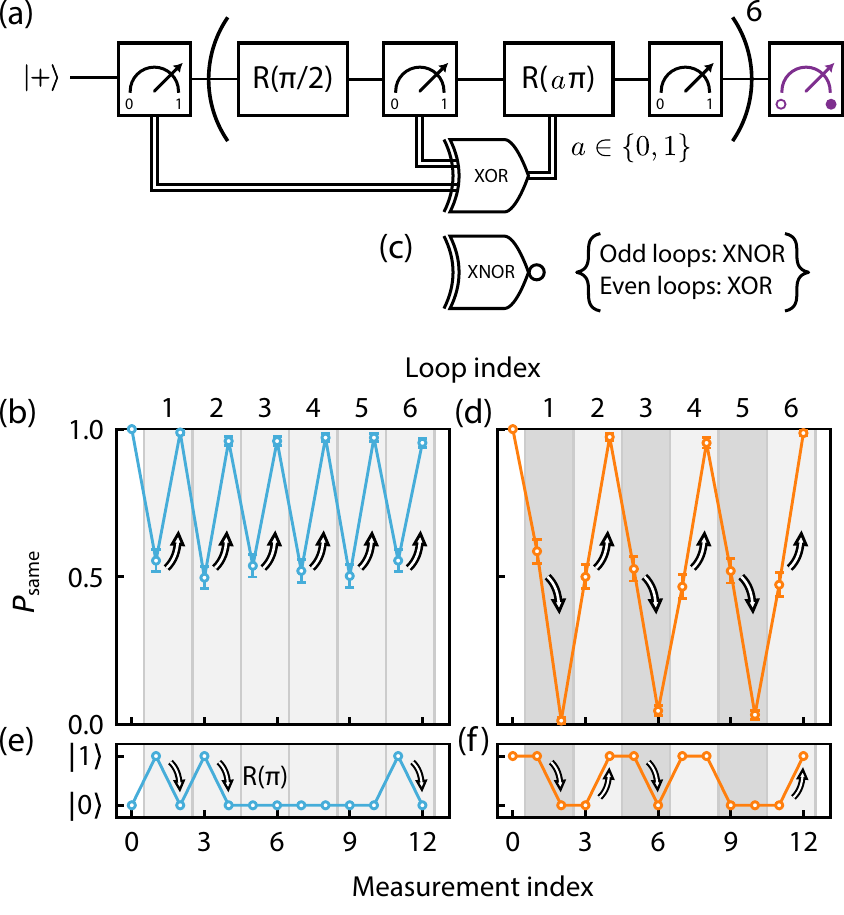}
    \caption{
        \textbf{Repetitive real-time feedforward}. (a) and (c) The circuit for performing repetitive deterministic state preparation for a single qubit. After initializing in $|+\rangle$ and performing an initial measurement, we repeat a loop that contains a first pulse ($\pi/2$), a first measurement, a second pulse ($0$ or $\pi$), and a second measurement. In (a), we choose the second pulse based on the first measurement outcome in order to yield the outcome of the second measurement to match the initial measurement outcome. This requires feeding forward the result obtained from an exclusive OR (XOR) gate between the classical bits in the initial measurement and the first measurement in the loop. The result is shown in (b), where the thick solid line shows the result averaged over $400$ shots. In (c), we choose to alternate between obtaining the opposite and the same outcome in the second measurement as that of the initial measurement, which requires alternating between exclusive not OR (XNOR) and XOR classical logic on odd- and even-numbered loops. The result is shown in (d), where now we see full-contrast alternations. The arrows in (b) and (d) illustrate the role of the feedforward, with gray vertical lines dividing loop iterations. Loops using XOR logic have light gray backgrounds and loops using XNOR logic have dark gray backgrounds. (e) and (f) show a single, representative trajectory for each circuit and the arrows indicate when a $\pi$-pulse is applied.
        \label{Figure7}
    }
\end{figure} 

\section{Real-time feedforward for active qubit reset}\label{feedforward}
We now add real-time control~\cite{Singh2022} to our toolbox to deterministically prepare the qubit in either $|0\rangle$ or $|1\rangle$ after measurement, often called ``active qubit reset"~\cite{Geerlings2013,Monz2016,Magnard2018,Chertkov2022}. We perform this study with only a single atom; a possible extension to arrays is discussed in Section~\ref{Outlook}. We initialize the qubit in $|+\rangle$ and then perform a measurement that projects it to $|0\rangle$ or $|1\rangle$. We then perform six loops, where each is composed of a $\pi/2$ rotation, a measurement, a conditional rotation ($R(0)$ or $R(\pi)$), and a second measurement. The goal of the conditional rotation is to rotate the qubit to the same or opposite state as measured in the initial readout. Hence, the rotation is conditional on both the initial readout and the first readout in the loop. Details on the real-time implementation of this circuit are described in Appendix~\ref{feedforward-app}. 

To keep the qubit in the same state as the outcome of the initial measurement [see Fig.~\ref{Figure7}(a) and (b)], a classical exclusive OR (XOR) gate is used to perform a $R(\pi)$ rotation if and only if the two classical bits are different. Otherwise, no rotation is performed: $R(0)$. We observe alternation between $P_\t{same}\approx1$ and $P_\t{same}\approx0.5$, where ``same'' refers to the initial readout. The first jump is identical to the $R(\pi/2)$ case of the Zeno study above [see Fig.~\ref{Figure6}(b)]; however, instead of staying at $P_\t{same}=0.5$ on average after each subsequent rotation, our feedforward technique deterministically puts the qubit back into the initial state, such that $P_\t{same}$ goes back to unity. The jump then repeats in each loop. 

Alternatively, we can switch between obtaining the initial readout outcome ($P_\t{same}=1$) and the opposite outcome ($P_\t{same}=0$) on each iteration of the loop. This can be accomplished by using an exclusive NOR (XNOR) in odd-numbered loops and an XOR in even-numbered loops [see Fig.~\ref{Figure7}(c)]. In this case, we observe full-amplitude zig-zags of $P_\t{same}$ as shown in Fig.~\ref{Figure7}(d). Individual trajectories are shown in Fig.~\ref{Figure7}(e) and (f), indicating when a $\pi$-pulse is required and showing ``quantum jump"-like behavior between sequential measurements.

Finally, since the qubit is deterministically reset, depolarization during readout does not accrue upon subsequent measurements. Thus, a high contrast is maintained for an arbitrary number of loops; we choose six loops (13 total measurements) as a compromise against atom loss. Note that this data is post-selected on the final atom readout.

\vspace{7mm}
\section{Concluding discussion}\label{Outlook}
In summary, we have leveraged the unique level structure of $^{171}$Yb to perform repetitive qubit readout, with qubits encoded in the nuclear spin-1/2 ground state. With a bright/dark discrimination fidelity of $\mathcal{F}\gtrsim0.99$, atom survival of $\approx0.99$, and spin-flip probability $\bar{P}_\t{depol}$ of $\approx 0.01$ during a \updated{$12\,\t{ms}$} probe pulse, we show that readout can be repeated 10 times while still maintaining control over the qubit state at the $\approx0.9$ level. These numbers would improve as $\sim1/B^2$. By adding high fidelity qubit rotations with an AC magnetic field, we study quantum circuits that feature both measurements and rotations. These include measurements in variable bases as a demonstration of measurement non-commutivity akin to cascaded Stern-Gerlach devices and optical polarizers, as well as a manifestation of the quantum Zeno mechanism in which dynamics is frozen by measurement. Additionally, we use real-time feedforward to perform repetitive active qubit reset to $|0\rangle$ or $|1\rangle$, and we show that this deterministic operation circumvents the accrual of depolarization errors during measurement thereby maintaining excellent contrast after 13 measurements, limited only by atom loss. The atom loss can be mitigated by using shallower tweezers and/or shorter readout pulses, or by working at a tweezer wavelength with larger detuning from the $^3$P$_J$$\leftrightarrow$ $^3$S$_1$ transitions for which we identify $\approx778$ nm as a good candidate where $^3$P$_1$ $|m_F|=3/2$ is magic with the ground state (see Appendix~\ref{polarizability}).       

Our work demonstrates global qubit rotations and global QND readout; however, local operations are often required. Local qubit rotations can be accomplished by using stimulated Raman pulses~\cite{Barnes2021,Jenkins2022,Chen2022} instead of an AC magnetic field. Qubit rotations can then be performed at $\sim$MHz rates rather than $\sim100$ Hz, and single-site control can be realized with adaptive optical elements~\cite{Barnes2021,Graham2022}. Local measurements, often referred to ``mid-circuit measurements", are performed on a subset of ``ancilla" qubits while the others -- the ``data'' qubits -- remain unaffected. Local mid-circuit measurement without crosstalk with the data qubits can be performed using two atomic species~\cite{Schmidt2005,Hume2007,Jiang2009, Nakajima2019,Xue2020,Singh2022}, separate readout zones~\cite{Wan2019,Pino2021,Dordevic2021,Bluvstein2022,Deist2022b}, and via ``shelving'' with other atomic states~\cite{Monz2016,Erhard2021,Graham2023}. Our technique offers QND readout for all three approaches, and is compatible with shelving techniques via the optical ``clock'' transition~\cite{Allcock2021,Yang2022,Wu2022,Chen2022}. Indeed, this is our primary motivation for operating at the clock-magic wavelength. In this approach, qubits will be encoded in the metastable $^3$P$_0$ nuclear spin, and optical pulses with a phase-stabilized laser~\cite{Endo2018,Li2022} will transfer the ancilla qubit(s) to the ground state for measurement. We also note that our technique is compatible with the use of local light shifts to perform mid-circuit measurements.

Finally, we note that having a qubit with excellent optical bright/dark discrimination is a key prerequisite for time bin remote entanglement generation~\cite{Pfaff2013,Bernien2013}. In this scheme, the qubit state becomes entangled with the state of a single photon in the temporal basis (early or late bin), and photons from the pair of atoms are coincident on a photonic Bell state analyzer that haralds the generation of a remote Bell pair between the atoms. Our work demonstrates that $^{171}$Yb is an excellent candidate for high-fidelity atom-photon entanglement. This could either be performed on the $^1$S$_0\leftrightarrow~^3$P$_1$ transition at 556 nm used in this work for short distance distributed or modular computing~\cite{Jiang2007,Monroe2014,Hucul2015}, or via the identical configuration on the ${}^3\t{P}_0 \leftrightarrow {}^3\t{D}_1$ transition at 1389 nm in the telecommunication wavelength band that is well suited for long distance networking~\cite{Covey2019b,Huie2021}. 

\section*{Acknowledgments}
We thank Saeed Pegahan, Mingkun Zhao, Ian Vetter, and Brett Merriman for contributions to the experimental system. We acknowledge Hannes Bernien, Adam Kaufman, Kaden Hazzard, and Danial Stack for stimulating discussions, and Bryce Gadway for critical reading of the manuscript. We acknowledge funding from the NSF QLCI for Hybrid Quantum Architectures and Networks (NSF award 2016136); the NSF PHY Division (NSF award 2112663); the NSF Quantum Interconnects Challenge for Transformational Advances in Quantum Systems (NSF award 2137642); the ONR Young Investigator Program (ONR award N00014-22-1-2311); the AFOSR Young Investigator Program (AFOSR award FA9550-23-1-0059); and the U.S. Department of Energy, Office of Science, National Quantum Information Science Research Centers.\\

%% Supplemental Material
%\vspace{15mm}
\setcounter{section}{0}
\twocolumngrid

\renewcommand\appendixname{APPENDIX}
\appendix
\renewcommand\thesection{\Alph{section}}
\renewcommand\thesubsection{\arabic{subsection}}

%\begin{center}
%\textbf{APPENDIX}
%\end{center}

%\renewcommand\thefigure{\thesection.\arabic{figure}}    
%\setcounter{figure}{0}  

\section{Overview of the experimental apparatus}\label{apparatus}

\subsection{Chamber, MOT, and imaging}
The experimental apparatus is inspired by Ref.~\cite{Dorscher2013} and comprises
two main sections, wherein hot ytterbium atom flux obtained from a single
AlfaVakuo dispenser is first cooled in two dimensions via a 2D magneto-optic
trap (MOT) and then transferred $\approx 40\,\t{cm}$ via a nearly resonant beam
through a differential pumping tube to load a full 3D MOT with $\sim 10^6$ atoms
over $500\,\t{ms}$. Both MOTs and the push beam are tuned to the ${}^1\t{S}_0
\leftrightarrow {}^1\t{P}_1$ transition ($\lambda = 399\,\t{nm}$, $\Gamma = 2
\pi \times 29\,\t{MHz}$) and are formed at the centers of glass cells. Once
loaded in the 3D MOT, the atoms are approximately $1\,\t{mK}$.

Next, another 3D MOT tuned to the ${}^1\t{S}_0 \leftrightarrow {}^3\t{P}_1$
($\lambda = 556\,\t{nm}$, $\Gamma = 2 \pi \times 182\,\t{kHz}$) transition is
turned on while the $399\,\t{nm}$ MOT beams are turned off and the magnetic
field gradient switched from $\approx 60\,\t{G/cm}$ to $\approx 10\,\t{G/cm}$
accordingly. Initially, this transition is power-broadened significantly by beam
intensities set to $\sim 10^4\,I_\t{sat}$ to ensure sufficient atom transfer
(roughly $50\%$) between the two MOTs. The atoms are then cooled further to
approximately $5\,\t{$\mu$K}$ by ramping the beam intensity down to
$0.6\,I_\t{sat}$ and detuning (relative to free space) from $\approx -20 \Gamma$
to $\approx -1.2 \Gamma$ over $30\,\t{ms}$. The MOT field gradient is then
increased to $\approx 14\,\t{G/cm}$ over $10\,\t{ms}$ to compress the atoms into
a volume roughly $150\,\t{$\mu$m}$ in diameter. We estimate that the compressed
MOT holds approximately $5 \times 10^5$ atoms at this stage.

The atoms are then loaded stochastically into optical tweezers at spacing
$7.8\,\t{$\mu$m}$, $1/e^2$ waist $670\,\t{nm}$ (radius), depth
$580\,\t{$\mu$K}$, and wavelength $\approx 760\,\t{nm}$. The tweezers are
generated by a single acousto-optic deflector leading into a $\t{NA} \gtrsim
0.6$ objective (Special Optics). The tweezer light is sourced from a M Squared
SolsTiS titanium-sapphire laser pumped by a M Squared Equinox. We choose the
tweezer wavelength to give magic trapping for the ground and excited state
manifolds of the ${}^1\t{S}_0 \leftrightarrow {}^3\t{P}_0$ optical clock
transition and near-magic trapping for the $|m_F| = 1/2$ states in the
${}^3\t{P}_1\, F = 3/2$ manifold. For an array of 5 tweezers, we require roughly
$35\,\t{mW}$ of optical power in the plane of the atoms. We estimate a loading
fraction of approximately \updated{$0.7$} (see Appendix \ref{analysis}). The
atoms are then cooled using the same $556\,\t{nm}$ beams used for the 3D MOT,
now with intensity $1.3\,I_\t{sat}$ and frequency red-detuned from the
free-space ${}^1\t{S}_0 \leftrightarrow {}^3\t{P}_1 \ket{F = 3/2,\, m_F = -1/2}$
transition by approximately $0.8\,\Gamma$, which causes atoms to escape from the
tweezers in pairs through light-assisted collisions, leaving only 0 or 1 atom in
the trap afterward. This process takes $\approx 120\,\t{ms}$, although we expect
this could be improved significantly with further optimization. We measure the
temperature of the atoms in the tweezers using a release-and-recapture method to
be $\approx 5\,\t{$\mu$K}$, which is close to the Doppler limit for the
transition.

The atoms are imaged using two retro-reflected beams tuned to the ${}^1\t{S}_0
\leftrightarrow {}^3\t{P}_1$ transition (see Appendix \ref{32vs12}) with
projections onto all three trap axes. The two beams are collimated with a
$1/e^2$ radius of approximately \updated{$880\,\t{$\mu$m}$} and have a $\approx
70^\circ$ angle between them. We estimate that each imaging beam has intensity
\updated{$0.5\,I_\t{sat}$} relative to the probe transition (other
polarization components are not counted). The two probe beams have polarization
overlap, and the polarizations of the retro-reflection beams are not rotated.
Therefore, interference fringes are likely to be present; we do not wash them
out with e.g. dithering mirrors. Imaging performed using either the $m_F = -3/2$
or $m_F = -1/2$ transition is done so with laser frequency red-detuned by
approximately \updated{$1 \Gamma$}. Atomic fluorescence is collected through a
second objective identical to the one used to generate the tweezers but placed
on the opposite side of the glass cell [see Fig. \ref{Figure1}(a)] and focused
onto an electron-multiplying CCD (EMCCD, Andor iXon Ultra 888) with EM gain set
to \updated{200}.

\subsection{Tweezer array homogenization}
Homogenization of the optical tweezer array is key to the maintenance of the
imaging condition used for readout. Although the atoms in the tweezers are
inherently identical, it is critical -- particularly given the nonzero
differential polarizabilities identified in Appendix \ref{polarizability} --
that the trapping potentials be as uniform as possible to prevent undesired
light shifts on the ${}^3\t{P}_1$ imaging states.

To this purpose, we adopt an iterative procedure based on spectroscopy of the
${}^3\t{S}_0 \leftrightarrow {}^3\t{P}_1\, F = 3/2,\, m_F = -3/2$ transition.
Since this transition is non-magic in the presence of the chosen trapping
wavelength, the measurement of the transition's resonance frequency is linearly
related to the trap depth. Thus, spectroscopy is repeatedly performed for each
site in the array, and the amplitudes of the five generating RF tones sent to
the acousto-optic deflector (AOD, AA Opto Electronic DTSX-400-760) from an
arbitrary waveform generator (AWG, Spectrum Instruments M4i6622) are adjusted to
bring the transition resonances measured across different sites to the same
center frequency. Then, the total RF power sent to the AOD is adjusted to bring
each site to the desired trap depth. For an array of five tweezers, this process
generally converges to the $\sim 0.1\%$ level in around 10 iterations.
Post-imaging atom survival is homogenized by this procedure to within $3\%$.
Uniformity in the shapes and depths of the tweezers is monitored by a CCD placed
after a dichroic mirror used to separate atomic fluorescence from tweezer light
after they have both passed through the imaging objective shown in
Fig.~\ref{Figure1}(a). We have observed similar homogeneity with 10 tweezers.

\subsection{Brief overview of experiment control} The many
individual components of the apparatus are controlled by means of a combination
of National Instruments PCIe-7820 and PCIe-6738, which respectively expose 128
digital input/output and 32 analog output-only configurable voltage channels,
housed in a single computer. Communication with these devices is accomplished by
means of low-level field-programmable gate array (FPGA) programming software
provided by Entangleware, Inc. Experimental sequences are ultimately programmed
through a high-level, Python-based interface. The atomic signal returned from
the apparatus via collection on the EMCCD sensor is sent to a separate computer,
which handles both real-time feedforward (see Appendix~\ref{feedforward-app})
and the AWG used to control the tweezer array.

\section{Comparison of $m_F = -3/2$ and $m_F = -1/2$ readout}\label{32vs12}
In this section, we compare aspects of the two methods of performing readout
(i.e. using the $m_F = -3/2$ and $m_F = -1/2$ ${}^3\t{P}_1$ $F = 3/2$ excited
state). As stated in the main text, the principal reason to prefer one to the
other is that the $m_F = -3/2$ method is state-selective due to dipole selection
rules and, hence, can be used to convert bright/dark classifications into qubit
state measurements.

The exact imaging conditions used in the two cases shown here differ by only the
strength of the magnetic field at which they are performed. We compare the
methods by exposing the atoms to light from the probe beams at total intensity
\updated{$\approx 3\,I_\t{sat}$}, of which we estimate $\approx
1\,I_\t{sat}$ effectively drives the desired transition, but $m_F = -3/2$
imaging is performed at $58\,\t{G}$ with detuning \updated{$-1 \Gamma$} for
\updated{$12\,\t{ms}$} while $m_F = -1/2$ imaging is performed at $18\,\t{G}$
with detuning $-0.5 \Gamma$ for $20\,\t{ms}$. The lower field strength is used
to minimize effects from the Zeeman splitting between the nuclear spin ground
states, and the longer probe time is because the data was taken before
final optimization of the collection efficiency. Fig. \ref{m3h-m1h-comp}(a)
shows that the mean numbers of photons collected from filled tweezer sites (see
Appendix \ref{analysis}) differs only slightly with comparable discrimination
fidelities. Fig. \ref{m3h-m1h-comp}(b) gives temperature estimates, obtained via
a standard release-recapture experiment, for both cases as well, showing
$\approx 5\,\t{$\mu$K}$ for $m_F = -3/2$ and $\approx 4\,\t{$\mu$K}$ for $m_F =
-1/2$. We do not explicitly measure the axial temperature and we expect that it
could be improved with an axial cooling beam sent through the objectives. We
leave this investigation for future studies.

\begin{figure}[t!]
    \includegraphics[width=\linewidth]{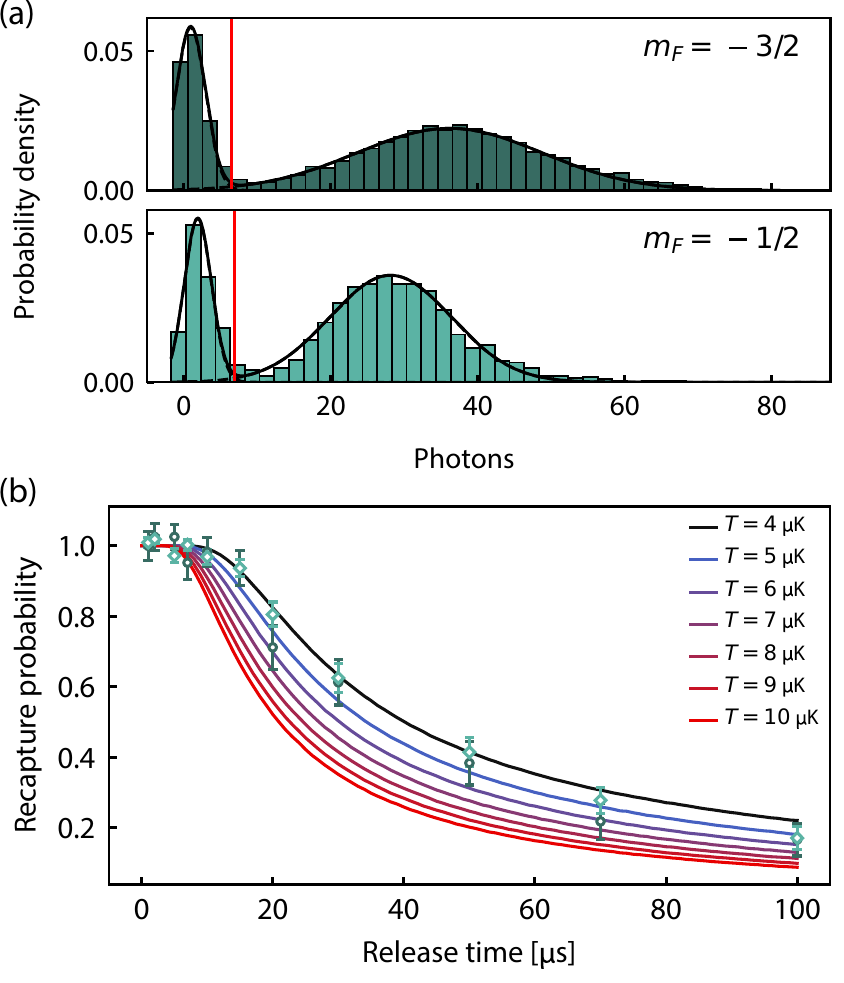}
    \caption{
        \textbf{Comparison of the $m_F = -3/2$ and $m_F = -1/2$ imaging
        conditions.} (a) Comparison of photon scatter during imaging. The upper
        histogram shows that for imaging using the $m_F = -3/2$ excited state
        with mean number of photons collected $\mu_1 = 37.1(2)$ at $58\,\t{G}$
        magnetic field when an atom is present (see Appendix \ref{analysis})
        while the lower shows that for imaging using $m_F = -1/2$ with $\mu_1 =
        28.1(2)$ at $18\,\t{G}$. The red vertical lines show the photon
        collection threshold for bright/dark discrimination in each case (see
        Appendix \ref{analysis}). (b) Measurement of atom temperature in the
        tweezer under the $m_F = -3/2$ (dark green circle) and $m_F = -1/2$
        (light green diamond) via release-recapture experiment. Tweezer sites
        are imaged and the tweezers are diabatically switched off for variable
        time before being turned on and imaged again. Recapture probability is
        calculated as that of the second image being bright, post-conditioned on
        the first being bright as well. Optical pumping is performed before both
        images in the $m_F = -3/2$ case. Probabilities have been re-scaled so
        that the mean of the four shortest-time probabilities is equal to 1. The
        lines show predicted probabilities obtained via Monte Carlo simulation
        for temperatures $4\,\t{$\mu$K}$ (black) up to $10\,\t{$\mu$K}$ (red),
        indicating that the temperature after imaging using $m_F = -3/2$ ($m_F =
        -1/2$) is approximately $5\,\t{$\mu$K}$ ($4\,\t{$\mu$K}$).
        \label{m3h-m1h-comp}
    }
\end{figure}

We also note differences in photon collection efficiency between the two
transitions. In the electric dipole approximation, the direction in which a
given photon will be radiated depends on the associated change in angular
momentum $\Delta m_F$ undergone by the atom, and is governed by the following
angular probability distributions:
\begin{equation}
    f_{|\Delta m_F|}(\theta, \varphi)
        = \begin{cases}
            \frac{3}{16 \pi} \Big( 1 + \cos^2\theta \Big)
                & |\Delta m_F| = 1
            \\
            \frac{3}{8 \pi} \sin^2\theta
                & |\Delta m_F| = 0
        \end{cases}
\end{equation}
with polar angle $\theta$ and azimuthal angle $\varphi$.

We then model the emission patterns for the two imaging cases detailed here. For
the $m_F = -3/2$ case, we expect the pattern to follow purely that of $f_1$
($|\Delta m_F| = 1$). For the $m_F = -1/2$ case, we expect it to follow a mix of
both $f_0$ ($|\Delta m_F| = 0$) and $f_1$ in a $2:1$ ratio equal to that of the
squared Clebsch-Gordan coefficients for the possible decay paths from the
${}^3\t{P}_1$ $F = 3/2$, $m_F = -1/2$ excited state, assuming an approximately
even mixture of polarization modes in the incident light. Thus the photon
emission for both imaging cases is modeled by the distributions
\begin{equation}
    g_{m_F}(\theta, \varphi)
        = \begin{cases}
            f_1(\theta, \varphi)
                & m_F = -3/2
            \\
            \frac{2}{3} f_0(\theta, \varphi) + \frac{1}{3} f_1(\theta, \varphi)
                & m_F = -1/2
        \end{cases}
.\end{equation}

From these, we then compare photon collection efficiencies between the
two cases via Monte-Carlo integration. Our objectives' collection areas are
modeled as circular, with bounding curves $\theta_\pm(\varphi) =
\arccos[\mp(\t{NA}^2 - \sin^2\varphi)^{1/2}]$, over which $g_{m_F}$ is
integrated for both $m_F = -3/2$ and $m_F = -1/2$. The ratio between the two
cases is found as approximately \updated{$1.4$}, in favor of $m_F = -1/2$, which
is in disagreement with our findings shown in Fig~\ref{m3h-m1h-comp}. We
attribute to a handful of factors. First, we note that the intensity
distribution in the probe beams across the $\pi,\, \sigma^\pm$ polarization
modes may not be uniform. The exact polarization of the beams is difficult to
measure given their angles of entry into the cell, and would manifest as
additional weighting factors on the $f_{\Delta m_F}$ components of the overall
photon emission distribution $g_{m_F}$. Second, interference between the
two probe beams may also play a role, as noted in
Appendix~\ref{apparatus}. Finally, the non-magic trapping of the $m_F =
-3/2$ excited state implies that the detuning of the probe beam varies spatially
over the trap, which causes broadening in the photon distribution. A temperature
of $T\approx5\,\t{$\mu$K}$ gives rise to a frequency spread of $\Delta
f\approx25\,\t{kHz}$ with a differential polarizability of $0.25(3)$, which is
non-negligible compared to the probe transition linewidth. However, the excess
broadening may suggest that our atoms are somewhat hotter in the axial
direction, to which release-recapture measurements are not sensitive. We note
that it is straightforward to add an axial cooling beam through the objective,
and we leave this study for future work.

\section{Magnetic field system}
\subsection{AC magnetic field system}\label{acfield}
To manipulate the nuclear spin states, we can either use the Raman transition via the electric dipole coupling~\cite{Barnes2021,Jenkins2022} or directly using the magnetic dipole coupling between nuclear states~\cite{Ma2022}. Here we introduce the second method implemented in our experiment.

For the $^{171}$Yb ground state ($6s^2 ~ {}^1\textrm{S}_0$), the nuclear spin $I=1/2$ gives the hyperfine structure $F=1/2$ with two nuclear spin states $\ket{0}\equiv\ket{m_F=-1/2}$ and $\ket{1}\equiv\ket{m_F=1/2}$. We apply a magnetic field along the $y$-axis [see Fig.~\ref{Figure1}(a)], leading to Larmor precession in the atomic spin. The Hamiltonian of the system is given by
\begin{equation}
    \hat{H_0}=g\mu_B \hat{F} \cdot B_0
,\end{equation}
where $g\mu_B/h$ is around $-750\,\t{Hz/G}$ for the splitting between $\ket{0}$ and $\ket{1}$. Thus the Hamiltonian can be simplified as:
\begin{equation}
    \hat{H_0}=\hbar\begin{pmatrix}\omega_0 & 0\\ 0 & 0\end{pmatrix}
,\end{equation}
where $\hbar\omega_0$ is the energy splitting between $\ket{0}$ and $\ket{1}$. Now considering we add an AC magnetic field along the $z$-axis, with the quantization axis defined by the magnetic field along the $y$-axis shown in Fig.~\ref{Figure1}(a), the Hamiltonian is given by:
\begin{equation}
\begin{split}
    \hat{H}
    &=\hat{H_0}+\frac{1}{2}\hbar\Omega \cos(\omega t)\hat{\sigma}_x \\
    &=\hbar\begin{pmatrix}\omega_0 & \frac{1}{2}\Omega \cos(\omega t)\\ \frac{1}{2}\Omega \cos(\omega t) & 0\end{pmatrix},
\end{split}
\end{equation}
where $\Omega=\frac{1}{2}g\mu_B B_\t{AC}/\hbar$ is the Rabi frequency (defined as the frequency of oscillation in state probabilities, rather than amplitudes), $B_\t{AC}$ is the strength of the driving magnetic field and $\omega$ is the frequency of the driving signal. Since $\omega_0\gg\Omega$, the factor $1/2$ in $\Omega$ comes from the strength of the counter-rotating term, which does not significantly contribute to the Rabi oscillation and, hence, has been neglected.

\begin{figure}[t!]
    \centering
    \includegraphics[width=0.49\textwidth]{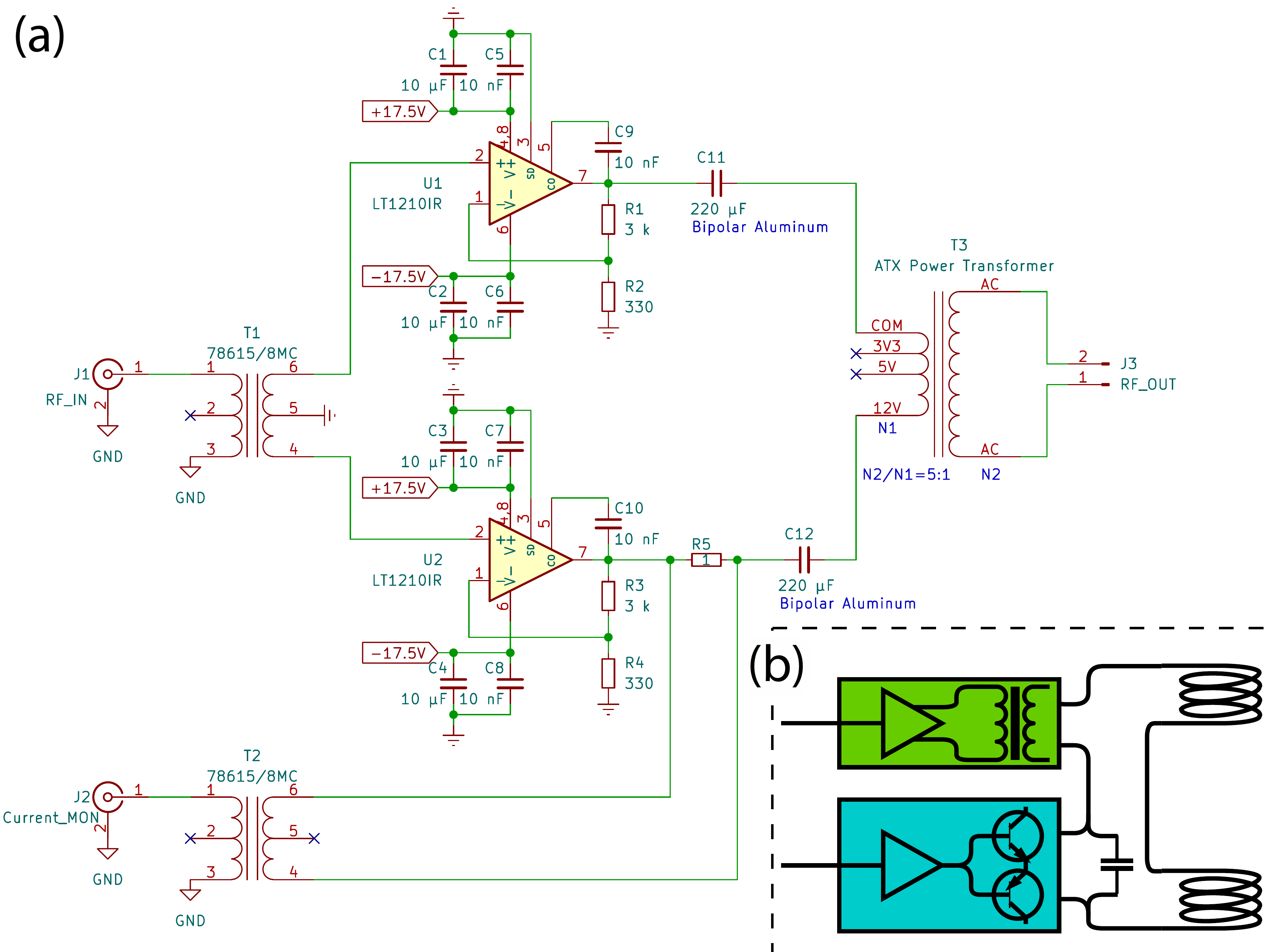}
    \caption{
        \textbf{AC magnetic field system} (a) The schematic of the RF driver used to drive one pair of the horizontal shim coils. The driver is capable of giving a maximum output voltage around $300\,\t{V\textsubscript{pp}}$ and the peak current about $0.2\,\t{A}$ ranges from $30\,\t{kHz}$ to $2\,\t{MHz}$. (b) The connection of the RF driver to the shim coils. The isolated output of the RF driver (green box) is connected in series with the shim coil driver (teal box). A $1\,\t{$\mu$F}$ polypropylene film capacitor is used to bypass the RF signal through the shim coil driver.
        \label{rfdriver}
    }
\end{figure}

To achieve a Rabi frequency $\Omega/2\pi \approx 110\,\t{Hz}$, we require $B_\t{AC} \approx 0.29\,\t{G}$. The easiest way of realizing such magnetic field modulation is by using a pair of ``shim'' coils [see Fig.~\ref{Figure1}(a)]. In our system this corresponds to a current modulation amplitude $I_\t{mod} \approx 0.15\,\t{A}$. At $58\,\t{G}$, chosen to match the magnetic field applied during readout, the modulation must be applied at a RF frequency equal to the Larmor frequency, $f \approx 43.5\,\t{kHz}$. Considering an inductance $L_\t{coil} \approx 1.5\,\t{mH}$ of the shim coil, the voltage modulation amplitude is calculated as $V_\t{mod} = I_\t{mod} \times 2 \pi f L_\t{coil} = 64\,\t{V}=128\,\t{V\textsubscript{pp}}$.

Since there is no commercial product that can give both a high voltage and a high current drive at $\approx45$ kHz, we built our own driver. The schematic of this RF driver is shown in Fig.~\ref{rfdriver}(a). The input RF signal is isolated and divided into two parts with opposite polarity by the signal transformer T\textsubscript{1}. Two high current and high speed operational amplifiers (Linear Technology, LT1210) are driven by the two signals and differentially drive the output transformer T\textsubscript{3}. T\textsubscript{3} is the main transformer recycled from an ATX (Advanced Technology eXtended) power supply of retired computers. This transformer has a turn ratio of about $5 : 1$ which increases the output voltage by a factor of 5. The output transformer also isolates the RF driving stage from the shim coils, which introduce minimum interference with the DC magnetic field control system. Since the maximum output current of LT1210 is $1.1\,\t{A}$, the maximum current available after the transformer is $0.2\,\t{A}$.

It is also necessary to use these shim coils with constant DC component in order to cancel background magnetic fields -- for which we also use two other pairs of coils not shown in Fig.~\ref{Figure1}(a) -- requiring the RF driver to be connected in series with a DC driver used for all three pairs. This is possible since the secondary winding of the transformer T\textsubscript{3} has very small DC resistance. However, the magnetic core of T\textsubscript{3} will be saturated if the DC current is larger than $0.3\,\t{A}$, which means the RF driver is only functional when the shim coils have relatively small DC current. This is not a problem for our experiment since the background magnetic field is small. As shown in Fig.~\ref{rfdriver}(b), a $1\,\t{$\mu$F}$ polypropylene film capacitor is connected in parallel with the shim coil driver, which is used to bypass the RF signal that passes through the shim coil driver. This capacitor also helps decrease the interference of the shim coil driver from the RF driver.

\subsection{DC magnetic field system}\label{dcfield}
To generate the magnetic field for non-destructive imaging at $58\,\t{G}$, we repurpose the MOT coils by switching the coils' electrical connection from anti-Helmholtz to Helmholtz configuration. To achieve this, we use an H-bridge to control the current flow direction of the upper MOT coil. The H-bridge is constructed using eight high-current metal-oxide-semiconductor field-effect transistors (MOSFETs) connected in parallel. The specific model used is IXFN300N20X3 from IXYS Corporation. The voltage-controlled resistors are used to protect each individual device from the back-electromotive force (EMF) generated by the coils. 

To isolate the noise and the computer's control signal ground from the coil's ground connection, two isolated gate drivers (Texas Instruments, UCC21320) are used to control both sides of the half bridges. The drivers also protect each side of the half bridge from the dead-zone, which can cause a short in the coil connection.

The magnetic field stabilization is achieved by stabilizing the current flowing through the coil. To measure the current with high stability, a high-stability Hall sensor (Danisense, DS300ID) is used, which has a long-term stability of better than $0.2\,\t{ppm}$ per month. The secondary current output of the Hall sensor is converted to voltage through a Kelvin connection, using a high stability ($0.05\,\t{ppm/\textsuperscript{$\circ$}C}$) metal film resistor (Vishay, Y16065R00000F9W). For the error signal amplifier, a low-noise operational amplifier (Analog Devices, AD8675) is DC stabilized by a zero-drift operational amplifier (Linear Technology, LTC2057), and high gain stability ($0.2\,\t{ppm/\textsuperscript{$\circ$}C}$) is achieved by a matched resistor network (Linear Technology, LT5400). The reference signal for the magnetic field servo is provided by a 20-bit high-stability ($0.05\,\t{ppm/\textsuperscript{$\circ$}C}$) digital-to-analog converter (Analog Devices, AD5791), with the reference voltage provided by a temperature-controlled buried Zener diode (Linear Technology, LTZ1000). The servo output is sent to two insulated gate bipolar transistors (IGBTs) connected in parallel, with each IGBT in series with the coils. The specific model used is IXGN200N170, manufactured by IXYS Corporation. Detailed schematics of the servo and voltage reference can be found in Ref.~\cite{Li2021}.

The magnetic field servo board and voltage reference board are connected closely inside a metal box that is grounded. A low noise isolated DC-DC power supply~\cite{Williams1997, Markell2000} is used to power the servo, voltage reference and the Hall sensor, each with a separated ground. The magnetic field control is achieved by the isolated digital channels that directly control the digital to analog converter using the serial peripheral interface (SPI) protocol. This approach avoids any possible magnetic field change introduced by the noise of external signal ground.

\section{Simulating multi-level dynamics}\label{multilevel}
During the qubit readout process, the relevant energy levels are those in the $^1$S$_0$ and $^3$P$_1$ $F=3/2$ manifolds. To estimate the readout fidelity and to optimize experimental settings, we study the dynamics of atom population transfer between the $^1$S$_0$ nuclear spin ground states via coherent and incoherent processes mediated by the $^3$P$_1$ $F=3/2$ states.

Since the $^3$P$_1$ state lifetime is relatively short compared with the imaging time ($\Gamma^{-1}\sim 10^{-6}$ s versus $\sim 10^{-2}$ s), we ignore the populations in $^3$P$_1$ states and only consider the populations in the two nuclear spin qubit levels in the $^{1}$S$_0$ manifold, $|m_F=-1/2\rangle\equiv|0\rangle$ and $|m_F=1/2\rangle\equiv|1\rangle$. As mentioned in Sec.~\ref{readout}, the qubit readout ($-3/2$ imaging) is achieved by applying a probe beam that is close to the $\ket{0}\leftrightarrow\ket{^3\mathrm{P}_1,m_F=-3/2}$ transition, red-detuned by $1\,\Gamma$. Due to the random polarization of the probe beams ($I_{\sigma^-}\approx I_{\sigma^+}\approx I_{\pi}=I/3$), off-resonant scattering processes via other $^3$P$_1$ states can induce qubit depolarization between $\ket{0}$ and $\ket{1}$. Moreover, the probe beam can coherently drive stimulated Raman transitions between $\ket{0}$ and $\ket{1}$ mediated by the $\ket{^3\mathrm{P}_1, m_F=\pm 1/2}$ states.

\begin{figure}[t!]
    \includegraphics[width=0.8\linewidth]{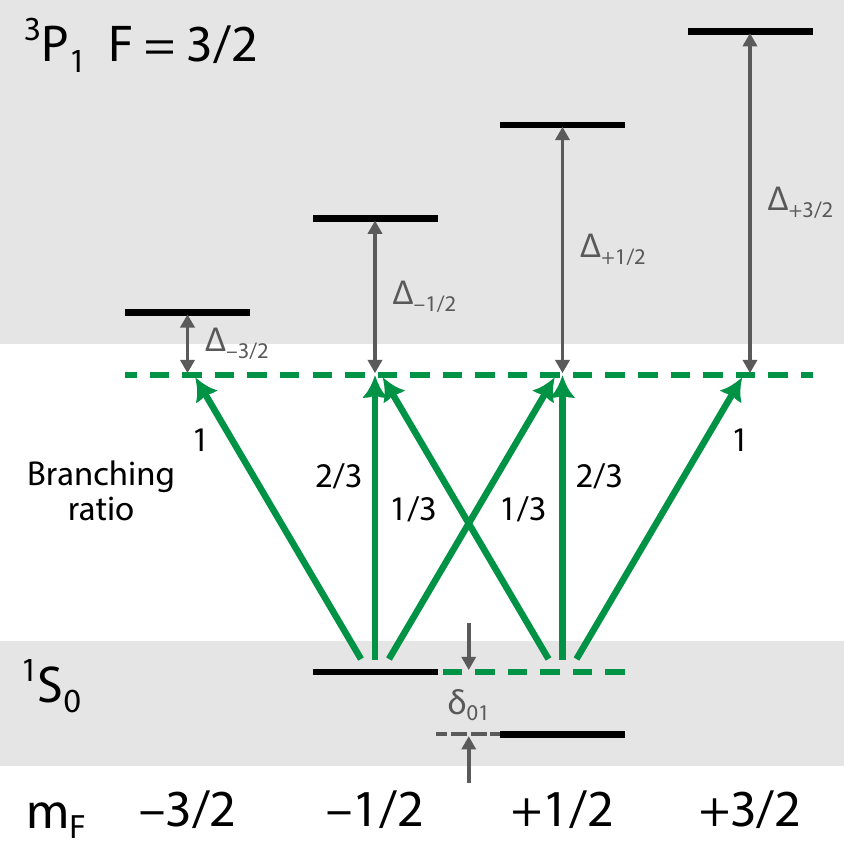}
    \caption{
    \textbf{Energy levels and probe beam configuration.} The energy splitting between different Zeeman states in $^1$S$_0$ and $^3$P$_1$ manifolds are determined by external magmetic field (Zeeman splitting) and tweezer depth (vector and tensor light shifts). For magnetic fields exceeding 5 G, the splitting between $^3$P$_1$ Zeeman states look nearly identical.
    }
    \label{EnergyLevels}
\end{figure}

Figure~\ref{EnergyLevels} shows the energy levels of $^1$S$_0$ and $^3$P$_1$ states and the probe beam with the corresponding branching ratio of each transition. Since the probe beam is monochromatic, the Hamiltonian of the stimulated Raman transition driven by the probe beam reads
\begin{align}
    \hat{H} = \frac{\hbar\Omega}{2}\hat{\sigma}_x+\frac{\hbar\delta_{01}}{2}\hat{\sigma}_z.
\end{align}
Here $\Omega=\Omega_{0}\Omega_{1}/2\Delta$ is the effective Raman Rabi frequency, with Rabi frequencies $\Omega_{0}$ and $\Omega_1$ of the dipole transition between a certain $^3$P$_1$ Zeeman state and the two ground states. Each of the two possible intermediate states, $\ket{^3\mathrm{P}_1,m_F=\pm 1/2}$, contributes to a Raman Rabi frequency of
\begin{align}
    \Omega_{\pm1/2} = \sqrt{\frac{1}{3}\frac{2}{3}}\frac{\Gamma^2}{2\Delta_{\pm1/2}}\frac{I/3}{2I_\t{sat}}
\end{align}
where $I_\t{sat}$ is the saturation intensity for the $\ket{0}\leftrightarrow\ket{^3P_1,m_F=-3/2}$ transition. The total Raman Rabi frequency is the sum of $\Omega_{-1/2}$ and $\Omega_{+1/2}$.

When the atoms are excited to the $^3$P$_1$ states, subsequent scattering leads to an incoherent re-distribution of the ground state populations, which can be described by the following collapse operators
\begin{align}
    \Gamma_{00} &= R_{00} \ketbra{0}{0}, & \Gamma_{01} &= R_{01} \ketbra{0}{1}, \nonumber
    \\
    \Gamma_{10} &= R_{10} \ketbra{1}{0}, & \Gamma_{11} &= R_{11} \ketbra{1}{1},
\end{align}
where $R_{ij}$ is the scattering rate from qubit state $\ket{i}$ to $\ket{j}$. Each scattering rate contains contributions from several paths via different Zeeman levels in the $^3$P$_1$ manifold that are allowed by the selection rule:
\begin{align}
    R_{00} &= R_{0 \rightarrow 0}^{-3/2} + R_{0 \rightarrow 0}^{-1/2} + R_{0 \rightarrow 0}^{+1/2}\nonumber\\
    R_{01} &= R_{0 \rightarrow 1}^{-1/2} + R_{0 \rightarrow 1}^{+1/2}\nonumber\\
    R_{10} &= R_{1 \rightarrow 0}^{-1/2} + R_{1 \rightarrow 0}^{+1/2}\nonumber\\
    R_{11} &= R_{1 \rightarrow 1}^{-1/2} + R_{1 \rightarrow 1}^{+1/2} + R_{1 \rightarrow 1}^{+3/2}.
\label{Eqn:ScatteringRate}
\end{align}
where $R_{i \rightarrow j}^m$ is the rate of the atom starting in $\ket{i}$ being excited to $\ket{^3\mathrm{P}_1, m_F=m}$ and decaying to $\ket{j}$. 
After taking the Clebsch-Gordan coefficients into consideration, the scattering rate via every possible channel in Eqs.~\ref{Eqn:ScatteringRate} can be calculated using rate equations. Equations~\ref{Eqn:scatter_T2} describe the transitions with the same initial and final state, representing a dephasing process,
\begin{align}
    R_{0 \rightarrow 0}^{-3/2} &= \frac{\Gamma}{2}\frac{I_{\sigma^-}/I_\t{sat}}{1+4(\Delta_{-3/2}/\Gamma)^2+I_{\sigma^-}/I_\t{sat}}\nonumber\\
    R_{0 \rightarrow 0}^{-1/2} &= \frac{\Gamma}{2}\frac{2}{3}\frac{2/3\cdot I_{\pi}/I_\t{sat}}{1+4(\Delta_{-1/2}/\Gamma)^2+2/3\cdot I_{\pi}/I_\t{sat}}\nonumber\\
    R_{0 \rightarrow 0}^{+1/2} &= \frac{\Gamma}{2}\frac{1}{3}\frac{1/3\cdot I_{\sigma^+}/I_\t{sat}}{1+4(\Delta_{+1/2}/\Gamma)^2+1/3\cdot I_{\sigma^+}/I_\t{sat}}\nonumber\\
    R_{1 \rightarrow 1}^{-1/2} &= \frac{\Gamma}{2}\frac{1}{3}\frac{1/3\cdot I_{\sigma^-}/I_\t{sat}}{1+4[(\Delta_{-1/2}+\delta_{01})/\Gamma]^2+1/3\cdot I_{\sigma^-}/I_\t{sat}}\nonumber\\
    R_{1 \rightarrow 1}^{+1/2} &= \frac{\Gamma}{2}\frac{2}{3}\frac{2/3\cdot I_{\pi}/I_\t{sat}}{1+4[(\Delta_{+1/2}+\delta_{01})/\Gamma]^2+2/3\cdot I_{\pi}/I_\t{sat}}\nonumber\\
    R_{1 \rightarrow 1}^{+3/2} &= \frac{\Gamma}{2}\frac{I_{\sigma^+}/I_\t{sat}}{1+4[(\Delta_{+3/2}+\delta_{01})/\Gamma]^2+I_{\sigma^+}/I_\t{sat}},
\label{Eqn:scatter_T2}
\end{align}
and Eqs.~\ref{Eqn:scatter_T1} describe the transitions between different initial and final states, which represent an amplitude damping process,
\begin{align}
    R_{0 \rightarrow 1}^{-1/2} &= \frac{\Gamma}{2}\frac{1}{3}\frac{2/3\cdot I_{\pi}/I_\t{sat}}{1+4(\Delta_{-1/2}/\Gamma)^2+2/3\cdot I_{\pi}/I_\t{sat}}\nonumber\\
    R_{0 \rightarrow 1}^{+1/2} &= \frac{\Gamma}{2}\frac{2}{3}\frac{1/3\cdot I_{\sigma^+}/I_\t{sat}}{1+4(\Delta_{+1/2}/\Gamma)^2+1/3\cdot I_{\sigma^+}/I_\t{sat}}\nonumber\\
    R_{1 \rightarrow 0}^{-1/2} &= \frac{\Gamma}{2}\frac{1}{3}\frac{2/3\cdot I_{\sigma^-}/I_\t{sat}}{1+4[(\Delta_{-1/2}+\delta_{01})/\Gamma]^2+2/3\cdot I_{\sigma^-}/I_\t{sat}}\nonumber\\
    R_{1 \rightarrow 0}^{+1/2} &= \frac{\Gamma}{2}\frac{2}{3}\frac{1/3\cdot I_{\pi}/I_\t{sat}}{1+4[(\Delta_{+1/2}+\delta_{01})/\Gamma]^2+1/3\cdot I_{\pi}/I_\t{sat}}.
\label{Eqn:scatter_T1}
\end{align}

With the above Hamiltonian and collapse operators, we can therefore use master equations to extract the simulation results in Fig.~\ref{Figure3}(b) for depolarization rates under different imaging conditions, including magnetic field, imaging time, probe beam intensity and detuning.

At high magnetic field, Fig.~\ref{Figure3}(b) indicates that the contributions from the coherent and incoherent parts are similar and they both scale as $1/B^2$. For the incoherent population transfer, the $1/B^2$ scaling can be explained by noticing that $R_{01}$ and $R_{10}$ both go as $1/B^2$ when $\Delta_{\pm 1/2} \gg \Gamma$ and $I\sim I_\t{sat}$. For the coherent part, the presence of all polarization components in our probe beam can drive stimulated Raman transitions between the qubit states, ostensibly at a rate of $~\Omega_{+1/2}+\Omega_{-1/2}\sim 1/B$. However, this rate is much smaller than the nuclear spin splitting $\delta_{01}$ at modest magnetic fields, meaning the coherent population oscillation is negligible. 

\section{Simulating the off-resonant scattering rate to $^3$P$_2$ and $^3$P$_0$}\label{atomloss}
At typical trapping power for the tweezer wavelength around $760\,\t{nm}$, the direct off-resonant scattering rates from the $^1$S$_0$ ground state to the $^3$P$_2$ and $^3$P$_0$ metastable states are negligible due to their narrow linewidth. However, there is still a possibility of the atoms being scattered to these two states through a two-photon process involving the $6s7s$ $^3$S$_1$ state when we probe the atoms on the $^1$S$_0\leftrightarrow ^3$P$_1$ transition.

To estimate this off-resonant scattering rate, we can simplify the calculation into two parts. First, we can calculate the probability of atoms being in the $^3$P$_1$ state during imaging. Second, we can calculate the off-resonant scattering rate from $^3$P$_1$ to the $^3$P$_2$ and $^3$P$_0$ states.

The probability of an atom occupying $^3$P$_1$ during the imaging process is given by the expression
\begin{equation}
\begin{split}
    P_{^3\textrm{P}_1}=\frac{1}{2}\frac{I_\t{probe}/I_\t{sat}}{1+4(\Delta/\Gamma_{^3\textrm{P}_1})^2+I_\t{probe}/I_\t{sat}}
\end{split}
\end{equation}
where $\Gamma_{^3\textrm{P}_1}$ and $I_\t{sat}$ represent the linewidth and saturation intensity of the probe transition, respectively. $I_\t{probe}$ and $\Delta$ denote the probe laser intensity and detuning, respectively.

Starting from $^3$P$_1$, the off-resonant scattering is dominated by the $^3$P$_1\leftrightarrow$ $ ^3$S$_1$ transition. For the case of large detuning and negligible saturation, the off-resonant scattering rate is written as:
\begin{equation}
\begin{split}
    \Gamma_{sc}=\frac{3\pi c^2 \Gamma^2_{^3\textrm{S}_1}}{2 \hbar \omega_0^3}\left(\frac{\omega}{\omega_0}\right)^3\left(\frac{1}{\omega_0-\omega}+\frac{1}{\omega_0+\omega}\right)^2 I_\t{trap},
\end{split}
\end{equation}
where $\Gamma_{^3\textrm{S}_1}$ and $\omega_0$ are the linewidth and resonance frequency of the transition. $\omega$ is the laser frequency and $I_\t{trap}$ is the intensity of the $760\,\t{nm}$ tweezer. To match the experiment, we set the tweezer power to be $7\,\t{mW}$ and the waist ($1/e^2$ radius) to be $670\,\t{nm}$.

However, the off-resonant scattering rate given above does not take into account the atom's initial $m_F$ state and the polarization of the tweezer, which results in a reduced scattering rate due to the reduction of dipole matrix elements. In our experiment, the tweezer is linearly polarized with the polarization parallel to the external magnetic field. Since $J=J'=1$, for the transition starting from the $\ket{F,m_F}$ lower state ($^3$P$_1$) to the $\ket{F',m_{F'}}$ upper state ($^3$S$_1$), the scattering rate is given by the expression 
\begin{align}   
\begin{aligned}
	\Gamma_{F,m_F}&=3\Gamma_{sc}\sum_{F',m_{F'}}\delta_{m_F,m_{F'}}(2F+1)(2F'+1)\\
	&\times
	\left|
	\begin{pmatrix}
		F' & 1 & F\\ 
		m_{F'} & -q & -m_F
	\end{pmatrix}
	\begin{Bmatrix}
		1 & 1 & 1\\ 
		F' & F & I 
	\end{Bmatrix}
	\right|^2,
\end{aligned}
\label{scatteringrate}
\end{align}
where $F'=1/2\textrm{ or }3/2$ for the $^3$S$_1$ state and $q=m_{F'}-m_F=0$ for a $\pi$-polarized tweezer. The symbols $(\,\cdots)$ and $\{\,\cdots\}$ represent Wigner 3-j and 6-j symbols, respectively. For our $|m_F|=1/2$ and $|m_F|=3/2$ imaging methods, the off-resonant scattering rates are given by $\Gamma_\t{sc}/6$ and $\Gamma_\t{sc}/2$, respectively.

\begin{figure}[t!]
    \centering
    \includegraphics[width=0.49\textwidth]{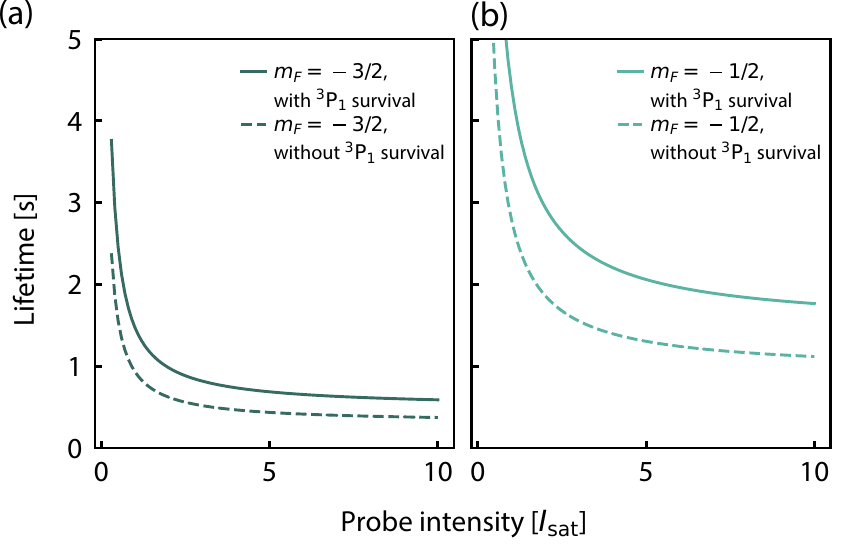}
    \caption{
        \textbf{Off-resonant scattering-limited probe lifetime} (a) The calculated lifetime under $|m_F|=3/2$ imaging as a function of probe intensity. The dashed lines assume the atom will be lost if it scatters to any $^3$P$_J$ state from ${}^3\t{S}_1$, while the solid lines assume it will survive only if scattered to $^3$P$_1$. (b) The lifetime under $|m_F|=1/2$ imaging for the same condition, which gives a three times higher lifetime due to the lower off-resonant scattering rate. We set the tweezer \{wavelength, power, waist ($1/e^2$ radius)\} to \{$760\,\t{nm}$, $7\,\t{mW}$, $670\,\t{nm}$\} for both calculations. A probe detuning of \updated{$-1\,\Gamma$} is used to calculate the $^3$P$_1$ state population. 
        \label{Lifetime}
    }
\end{figure}

After being off-resonantly excited to the $^3$S$_1$ state, the atom can decay back to any of the $^3$P$_J$ states. Since the $^3$P$_2$ state cannot be trapped by the $760\,\t{nm}$ tweezer, atoms in this state will leave the trap immediately. The $^3$P$_0$ state is dark to the probe transition and is not re-pumped in our apparatus, so both processes contribute to atom loss during the fluorescence imaging. By comparing the branching ratios of all three paths~\cite{Porsev1999}, we can calculate the possibility of decaying to the $^3$P$_2$ and $^3$P$_0$ to be around $63\%$ -- dominated by the $^3$P$_2$ component. The solid curves in Fig.~\ref{Lifetime}(a) and (b) show the calculation of the $|m_F|=3/2$ and $|m_F|=1/2$ imaging lifetime limited by the rate at which atoms decay to either the $^3$P$_2$ or $^3$P$_0$ states, assuming the atom survives if it decays to $^3$P$_1$. 

However, we note that, even if the atom decays to $^3$P$_1$, is not clear whether the atom survives after being pumped to the $^3$S$_1$ state for $\approx100\,\t{ns}$. Although the $^3$S$_1$ state is trappable with a tweezer wavelength of $760\,\t{nm}$, the trap depth is around 70 times higher than that of both $^3$P$_1$ and the ground state under the same tweezer power, which could introduce significant atom loss due to the sudden increase in potential energy. For comparison, we also plot the case that the atom is fully lost after being pumped to the $^3$S$_1$ state, as shown by the dashed curves in Fig.~\ref{Lifetime}(a) and (b). For the $|m_F|=3/2$ probe, this calculation gives an imaging lifetime of \updated{$\approx0.8\,\t{s}$} under typical experimental conditions ($7\,\t{mW}$ tweezer power, \updated{$1\,I_\t{sat}$ effective probe beam intensity with $-1\,\Gamma$ detuning}), which is in rough agreement with observation. For the $|m_F|=1/2$ probe, this calculation gives a lifetime of \updated{$\approx2.5\,\t{s}$}, which is longer than observation; we find similar probe lifetimes and survivals in both cases. It is not clear what other decay mechanism is in play. We measure a lifetime of $\tausurvvac$ for atoms in tweezers that are not illuminated by any light, which is limited by a combination of background gas collisions due to finite vacuum pressure as well as atomic heating due to intensity noise on the tweezer.

We also note that if an atom is off-resonantly excited to the $^3$S$_1$ state but eventually decays back to the original $^3$P$_1$ state and survives, the atom can still be depolarized between the $m_F=-1/2$ and $1/2$ ground states. Using a similar calculation as the one introduced in the Eq.~\ref{scatteringrate}, we can determine the probability of the depolarization process for the $|m_F|=3/2$ readout. After being off-resonantly pumped to the $^3$P$_1$ state, this probability is calculated to be $0.082$, considering decay from ${}^3\t{S}_1$ to ${}^3\t{P}_1$. This depolarization rate is significantly smaller than the off-resonant excitation rate, but would ultimately limit the depolarization during readout in situations where state mixing and multi-level dynamics can be neglected at a much higher magnetic field.  

\section{Polarizability calculations}\label{polarizability}
The potential that an atom experiences in an optical trap is given by the product of the state-dependent polarizability $\alpha$ and the spatially varying intensity profile $I(\mathbf{r})$ of the trap
\begin{equation}
	U_\mathrm{trap}(\mathbf{r}) = \frac{\alpha(\omega)}{2\epsilon_0 c}I(\mathbf{r}),
\end{equation}
where $\epsilon_0$ is the vacuum permittivity, and $c$ is the speed of light in vacuum. The dependence on the frequency of the trap laser and the atomic state in the polarizability can be understood from a quantum mechanical treatment of the induced dipole interaction energy, also known as the AC Stark shift. Following the derivations in Refs.~\cite{Steck2007,Kien2013}, the polarizability operator for an atomic state $\ket{i}$ can be written as a Cartesian tensor of the form
\begin{equation}
    \hat{\alpha}_{\mu\nu}(\omega) = \sum_{k} \frac{2\omega_{ki}}{\hbar(\omega_{ki}^2 - \omega^2)}\hat{d}_\mu\ketbra{k}{k}\hat{d}_\nu,
\end{equation}
where the sum is over all states connected to $\ket{i}$ via a dipole transition, $d_\mu$ is the projection of the dipole operator along the $\mu$-th component of the incident electric field, and $\omega_{ki}$ is the energy difference between the state $\ket{k}$ and $\ket{i}$. It is more insightful to decompose the $\alpha_{\mu\nu}$ into spherical components. The final result is given as
\begin{equation}
	\begin{aligned}
        \hat{\alpha}(\omega)=\alpha_S(\omega) - i\alpha_V(\omega)\frac{(\hat{\mathbf{u}}^*\times \hat{\mathbf{u}})\cdot \hat{\mathbf{F}}}{2F} \\
        +\alpha_T(\omega)\frac{3\{\hat{\mathbf{u}}^* \cdot \hat{\mathbf{F}},\hat{\mathbf{u}} \cdot \hat{\mathbf{F}}\} - 2\hat{\mathbf{F}}^2}{2F(2F-1)}.
	\end{aligned} \label{total polarizability}
\end{equation} 
The object $\{\cdot, \cdot\}$ is the anti-commutator of two operators. The coefficients $\alpha_S$, $\alpha_V$, and $\alpha_T$ are the scalar, vector, and tensor polarizabilities, respectively, of the atom for a given hyperfine state characterized by quantum numbers $\ket{nJIFm_F}$. The expressions for the individual polarizabilities are
\begin{align}
    \alpha_S &= \frac{1}{\sqrt{3(2J+1)}}\alpha^{(0)}_{nJ}\\
    \alpha_V &= (-1)^{J+I+F}\sqrt{\frac{2F(2F+1)}{F+1}}\begin{Bmatrix}
		F & 1 & F \\ J & I & J
	\end{Bmatrix} \alpha^{(1)}_{nJ}\\
    \alpha_T &= (-1)^{J+I+F +1}\sqrt{\frac{2F(2F-1)(2F+1)}{3(F+1)(2F+3)}}\begin{Bmatrix}
		F & 2 & F \\ J & I & J
	\end{Bmatrix} \alpha^{(2)}_{nJ}
\end{align}
where the reduced polarizabilities $\alpha^{(K)}_{nJ}$ are given by
\begin{align}
	\begin{aligned}
		\alpha^{(K)}_{nJ}&=(-1)^{K+J+1}\sqrt{2K+1}\\
		&\quad \times \sum_{n'J'}(-1)^{J'}\begin{Bmatrix}
			1 & K & 1 \\ J & J' & J
		\end{Bmatrix} \abs{\bra{n'J'}|d|\ket{nJ}}^2\\
		&\quad \times \frac{1}{\hbar}\left(\frac{1}{\omega_{n'J'nJ}-\omega}+\frac{(-1)^K}{\omega_{n'J'nJ}+\omega}\right).
	\end{aligned}
\end{align}
The reduced dipole matrix elements $\abs{\bra{n'J'}|d|\ket{nJ}}$ can be calculated from experimentally determined lifetimes of the relevant states via 
\begin{equation}
	\Gamma_{n'J'nJ}=\frac{\omega_{n'J'nJ}^3}{3\pi\varepsilon_0 \hbar c^3} \frac{\abs{\bra{J'}|d|\ket{J}}^2}{2J'+1}.
\end{equation}
A branching ratio will be needed if the excited state decays to several states of lower energy, such as in the case of decay from $^3$S$_1$ to $^3$P$_J$ that leads to atomic loss from the tweezer.

From \eqref{total polarizability}, we see that the tensor light shift vanishes for states with $F=0,1/2$. Moreover, the vector light shift vanishes when the tweezer is linearly polarized. We work mainly with a linearly polarized tweezer, hence the total light shift experienced by an atomic state is
\begin{equation}
	U_\mathrm{trap} = -\frac{I}{2\epsilon_0 c}\left(\alpha_S + \alpha_T\frac{3\cos^2\theta -1}{2}\frac{3m_F^2-F(F+1)}{F(2F-1)} \right) \label{light shift}
\end{equation}
where $\theta$ is the angle between the polarization of the tweezer and the axis of quantization set by an applied magnetic field. In the case of $^{171}$Yb, the $^1$S$_0$ and $^3$P$_0$ states have $F=1/2$, hence they only have a scalar contribution to the polarizability. On the other hand, the $^3$P$_1$, $F=3/2$ state will have all three components, in general.

\begin{figure}[t!]
	\centering
	\includegraphics[width=0.45\textwidth]{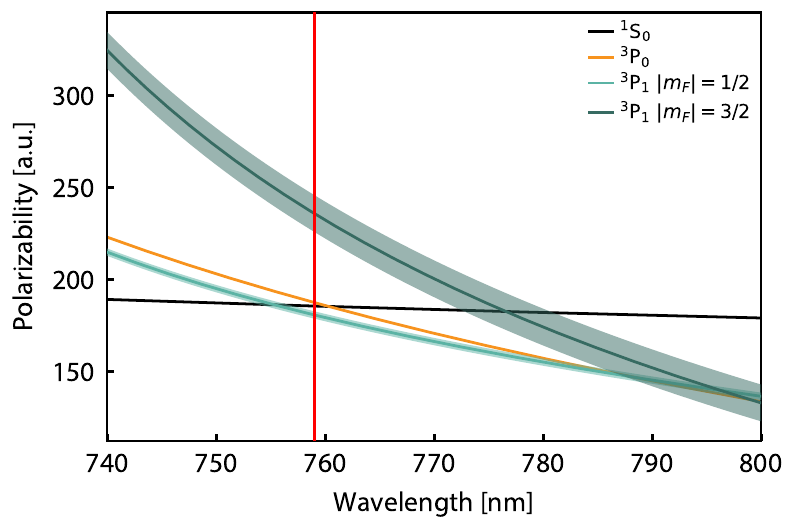}
    \caption{\textbf{Polarizabilities of the atomic states of interest.} The polarizabilities of the $^1$S$_0$ (black); $^3$P$_0$ (yellow); $^3$P$_1$, $F=3/2$, $|m_F| =1/2$ (light green); and $^3$P$_1$, $F=3/2$, $|m_F| = 3/2$ (dark green) are plotted as a function of the tweezer wavelength. The shaded band represents the uncertainty in the total polarizability, which arises from uncertainty in the trap waist and measured light shifts of the excited states. The magic wavelength ($759\,\t{nm}$, red vertical line) for the ground and clock states is also near-magic for the $|m_F|=1/2$ state. We have included correction factors in the calculations for the $^3$P$_1$ states to match the experimentally observed differential polarizabilities.}
	\label{polarizabilities}
\end{figure}

To calculate the polarizability of a given state, we perform a sum over states using measured values of the energy levels and lifetimes wherever possible. For the $^3$P$_1$ state, we use the reduced dipole matrix elements given in Ref.~\cite{Tang2018}. Our calculations yield $\alpha_S(^1\text{S}_0) = \alpha_S(^3\text{P}_0) = 186$, $\alpha_S(^3\text{P}_1) = 233$ au, and $\alpha_T(^3\text{P}_1) = 87$ au. The value for $\alpha_S(^1\text{S}_0)$ at the clock-magic wavelength of 759 nm is in excellent agreement with literature~\cite{Dzuba2010,Lemke2009}. This results in a predicted differential polarizability of $\alpha_{|1/2|}\approx-0.22$ and $\alpha_{|3/2|}\approx0.72$, which is significantly larger than the experimentally measured values of $\alpha_{|1/2|}\approx-0.030(3)$ and $\alpha_{|3/2|}\approx0.25(3)$ (see Sec \ref{readout}). As a result, we phenomenologically correct our calculated values by using Eq.~\eqref{light shift} to generate a set of linear equations
\begin{align}
	&\hbar\Delta\omega(|m_F|=3/2) \nonumber\\
	&\quad\quad =-\frac{I}{2\epsilon_0 c}\left(\alpha_S(^3\text{P}_1)+\alpha_T(^3\text{P}_1)-\alpha_S(^1\text{S}_0)\right) \\
	&\hbar\Delta\omega(|m_F|=1/2) \nonumber\\
	&\quad\quad =-\frac{I}{2\epsilon_0 c}\left(\alpha_S(^3\text{P}_1)-\alpha_T(^3\text{P}_1)-\alpha_S(^1\text{S}_0)\right)
\end{align}
which yields $\alpha_T(^3\text{P}_1) = 26(6)\,\t{a.u.}$ and $\alpha_S(^3\text{P}_1)-\alpha_S(^1\text{S}_0) = 20(4)\,\t{a.u.}$ Note that we operate at $\approx760.2$ nm, for which $\alpha_S(^1\text{S}_0)$ differs from the clock-magic wavelength by $\approx0.1$\%. The uncertainty in the values are mainly derived from the uncertainty in the measured differential polarizabilities and the uncertainty in the measured beam waist of the tweezer. We have ascribed a conservative estimate for the uncertainties of 10\%. Since the trap depth depends on the beam waist as $1/w_0^2$, the overall uncertainty is mainly dominated by the uncertainty in the beam waist. 

Using the corrected values, we plot the polarizability as a function of the tweezer wavelength in Fig.~\ref{polarizabilities}. It is interesting to note that the polarizabilities of all $^3$P$_1$ Zeeman states converge at $\approx 796\,\t{nm}$. Also, we identify $\approx778\,\t{nm}$ as a good candidate for implementing our readout technique because the $^3$P$_1$ $|m_F|=3/2$ states are magic with the ground state.

\section{Effect of vector and tensor light shifts}\label{mixing}
The vector and tensor light shift can cause additional state-mixing in the $^3$P$_1$, $F=3/2$ manifold, which opens up a depolarization channel when probing via the $m_F=-3/2$ state that causes the atoms to decay into the dark $^1$S$_0$, $m_F=1/2$ ($|1\rangle$) state. We estimate the level of state-mixing by numerically diagonalizing the AC Stark Hamiltonian together with the Zeeman Hamiltonian, and observe the complex amplitudes of the various eigenstates. 

We use a different convention for the coordinate system in Fig \ref{Figure1} for the following calculations. The applied magnetic field, which defines the axis of quantization, is oriented along the $+z$-axis, and the tweezer propagates along the $+y$-axis. Thus, the following parametrization for the polarization of the tweezer is valid:
\begin{equation}
\begin{aligned}
		\hat{\mathbf{u}} &= \left(\cos\gamma \cos\theta - i\sin\gamma\sin\theta \right)\hat{\mathbf{z}}\\
	&\quad + \left(\cos\gamma \sin\theta + i\sin\gamma\cos\theta \right)\hat{\mathbf{x}}
\end{aligned}
\end{equation}
where $0\leq \gamma \leq \pi/2$ represents the degree of ellipticity of the tweezer, and $\theta$ is the angle between the linear polarization of the tweezer and the axis of quantization set by the magnetic field. This parametrization allows us to consider cases where the tweezer is simultaneously rotated away from the axis of quantization and contains some degree of ellipticity. However, we will analyze the two cases separately by setting the counterpart to zero.

\begin{figure}[t!]
	\centering
	\includegraphics{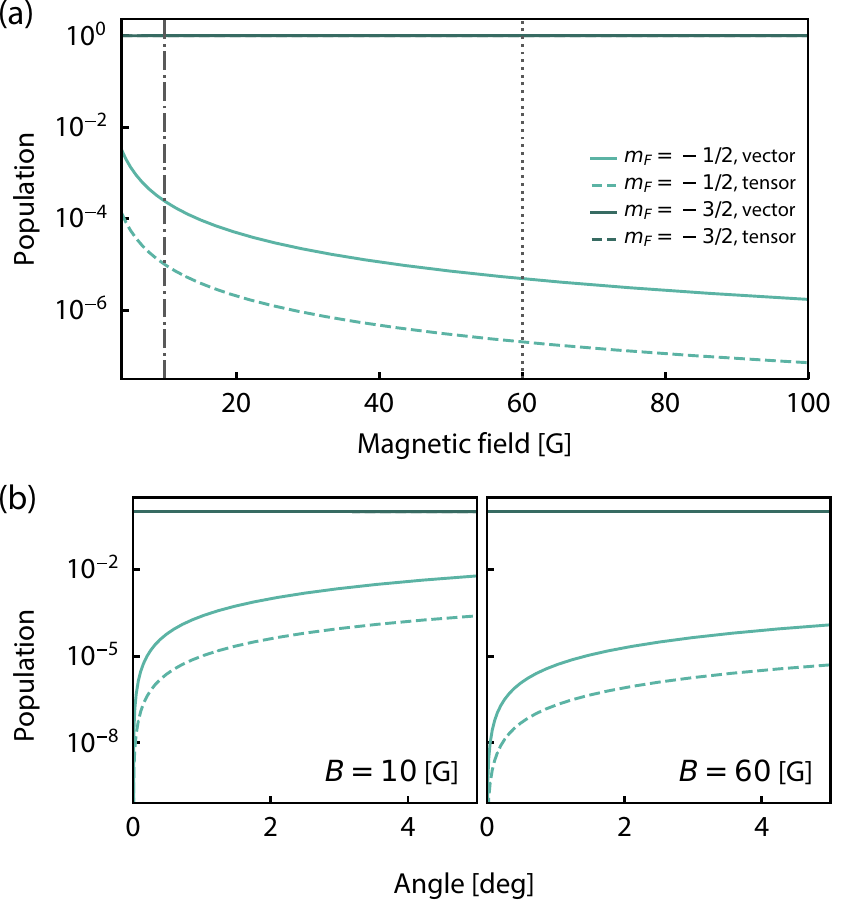}
    \caption{\textbf{State-mixing of the $^3$P$_1$, $F=3/2$, $m_F=-3/2$ imaging state due to the vector and tensor light shifts} (a) The probabilities in the $m_F=-3/2$ (dark green) and $m_F=-1/2$ states (light green) due to the vector ($\gamma=1$ degree and $\theta=0$ degrees; solid) and tensor ($\theta=1$ degree and $\gamma=0$ degrees; dashed) light shifts as the magnitude of the applied magnetic field is varied. The gray vertical lines at $10\,\t{G}$ (dot-dash) and $60\,\t{G}$ (dotted) show field values chosen for the angle studies shown in (b). (b) We study the dependence on the angles $\theta$ and $\gamma$, which describes the misalignment of the tweezer from the axis of quantization and the degree of ellipticity of the tweezer polarization, respectively. There is a sharp rise at angles close to zero degree due to the sudden appearance of the off-diagonal matrix elements.}  
	\label{statemixing}
\end{figure}

For the $^3$P$_1$, $F=3/2$, $m_F=-3/2$ state, the dominant state that the vector and tensor light shift mixes with is the neighboring $m_F=-1/2$ state. In Fig.~\ref{statemixing}(a), we consider the state-mixing as a function of the magnitude of the applied magnetic field. We consider $\theta$ or $\gamma = 1 ~\mathrm{deg}$ to illustrate the cases where the tweezer is not purely linear, or when the tweezer is indeed linear but not perfectly aligned onto the axis of quantization. At low fields, state-mixing is significant as the light shifts ($\sim1$ MHz) are comparable to the Zeeman energy ($\sim1.4 \times m_F$ MHz/G). The mixing can be suppressed by increasing the magnetic field. A similar trend can be seen when we fix the magnitude of the applied magnetic field and vary the angles. Whenever the angles are non-zero, the state-mixing is turned on as there is a competition between the axis of quantization defined by the applied magnetic field and those defined by the tweezer polarization. Naturally, it follows that as the applied magnetic field increases in strength, the effect of state-mixing decreases, which is confirmed by the results in Figure~\ref{statemixing}(b).The trend phenomenologically matches a $\sim1/B^2$ scaling.

For a scattering rate of \updated{$\approx97000\,\t{photons/s}$} and a probe time of \updated{$12\,\t{ms}$}, we estimate that \updated{$\approx1200$} photons are scattered by the atoms, which is consistent with our measured collection of \updated{$\approx36$} photons with a measured atom-to-camera collection efficiency of \updated{$0.04$}. For a depolarization probability of $\bar{P}_\t{depol}\approx0.01$ during the readout, we conclude that the depolarization probability per photon scatter is \updated{$\mathcal{P}_\t{depol} \lesssim 1 \times 10^{-5}$}. Up to a Clebsch-Gordan coefficient, we expect that this depolarization probability per photon is exactly equal to the contribution of the $m_F=-1/2$ state to the mixed eigenstate. As shown in Fig.~\ref{EnergyLevels}, the Clebsch-Gordan coefficients to decay from $^3$P$_1$, $m_F=-1/2$ favor $^1$S$_0$, $m_F=-1/2$ over $m_F=1/2$ by a factor of 2, so we can tolerate twice as much population in $^3$P$_1$, $m_F=-1/2$ than if the branching ratios were even for a given $\bar{P}_\t{depol}$. Therefore, this analysis suggests that we can tolerate a population of \updated{$\approx2.0\times10^{-5}$} in $^3$P$_1$, $m_F=-1/2$ to obtain $\bar{P}_\t{depol}\approx0.01$ during the readout, which for $B=58$ G corresponds to $\gamma$ or $\theta\approx1-2$ degrees. This estimate is consistent with experimental observations in which the tweezer polarization was deterministically moved on the Poincar\'e sphere using a polarimeter, and corroborates our observation of larger depolarization probabilities for tweezers at the edges of the array. 

\section{Characterization of state preparation and measurement parameters}
\label{analysis}
In this appendix, we describe several experimentally measured parameters
relevant to state preparation and measurement (SPAM) correction, the procedure
of which is described in Appendix~\ref{spam}. These are the tweezer loading
probability $p$, the bright- and dark-state readout survival probabilities
$\eta_\t{surv}^\t{B}$ and $\eta_\t{surv}^\t{D}$, the qubit spin-flip
probabilities $P_\t{depol}^\t{B\,$\rightarrow$\,D}$ and
$P_\t{depol}^\t{D\,$\rightarrow$\,B}$, and the $\pi$-pulse probability
$\eta_\pi$. We additionally define bright and dark discrimination fidelities
$F_1$ and $F_0$ which are also relevant to SPAM correction, but themselves left
uncorrected (see Appendix~\ref{spam}).

\subsection{Definition of base discrimination fidelity}
\label{base-fidelity}
Here we characterize the possible measurement-based error channels in our atom-
and state-readout detection schemes. All such errors derive from a common
source, which is the degree to which the data from a single camera exposure
while a tweezer site is illuminated can be correctly classified (or not) as
holding an atom in a fluorescent state. While it is straightforward to scatter
photons from a single atom in a tweezer and count the number detected by a
sensor, this measurement may be confounded by a number of other processes. For
example, it is possible that the atom is not in a state excited by a particular
laser frequency, the scattered photons may not all be collected by the sensor,
the atom may have exited the trap or gone dark during the exposure, or the atom
simply may not have been loaded in the first place.

We first define a base discrimination fidelity $\mathcal{F}$ as the probability
of correctly classifying a single image as containing a fluorescent atom, based
on the number of collected photons $x$. We assume a simple Gaussian mixture
model of two components governed by an overall distribution of the form
\begin{equation}
    \label{mixture-model}
    \Pi(x)
        = P_0 f_0(x) + P_1 f_1(x)
\end{equation}
where the mixture weights $P_N$ are the probabilities of $N$ such atoms in a
single tweezer, and $f_N$ is a Gaussian distribution with mean $\mu_N$ and
variance $\sigma_N^2$. Individual images are classified as ``bright'' ($B$,
indicating $N = 1$ fluorescent atom) if the total number of photons collected is
greater than or equal to some predetermined threshold value $\theta$, and
``dark'' ($D$, indicating $N = 0$ fluorescent atoms) otherwise. We therefore
define $\mathcal{F}$ more concretely as
\begin{equation}
    \label{fidelity-def}
    \begin{aligned}
        \mathcal{F}
            &= \t{Pr}(D | N = 0) \t{Pr}(N = 0)
            \\&\phantom{=}
            + \t{Pr}(B | N = 1) \t{Pr}(N = 1)
            \\
            &= P_0 \int_{-\infty}^\theta \dd{x} f_0(x)
            + P_1 \int_{\theta}^\infty \dd{x} f_1(x)
    \end{aligned}
\end{equation}
All the parameters of $\Pi$ are readily obtained by fitting to a measured
histogram of photon counts from a series of images, and $\theta$ is chosen to
maximize $\mathcal{F}$ at the number of photons where $P_0 f_0$ and $P_1 f_1$
intersect. We also define discrimination fidelities based on the individual
components of the mixture for use in SPAM correction,
\begin{equation}
    \begin{aligned}
        F_0
            &= \t{Pr}(D | N = 0)
            = \int_{-\infty}^\theta \dd{x} f_0(x)
        \\
        F_1
            &= \t{Pr}(B | N = 1)
            = \int_\theta^\infty \dd{x} f_1(x)
    .\end{aligned}
\end{equation}

In a typical experiment where we initialize atoms in tweezers by loading and
optically pumping to the bright state before imaging, we usually measure
\updated{$P_0 \approx 30\%$, and $P_1 \approx 70\%$}, with \updated{$\mu_0
\approx 1.0$, $\sigma_0 \approx 2$, $\mu_1 \approx 37$, $\sigma_1 \approx 16$}
photons. Note that we calculate photon counts using a linear transformation from
raw photoelectron counts yielded by our EMCCD (see Appendix \ref{apparatus}).
This transformation is calibrated to give zero photon counts at some non-zero
number of photoelectrons which, in relatively rare cases, can lead to negative
photon counts.

To estimate the uncertainty in $\mathcal{F}$ under both imaging conditions
described in Appendix \ref{32vs12} (treated independently), we use a
bootstrapping procedure following Ref.~\cite{Holland2022}. From a data set of
\updated{$2500 \times 5$} realizations of the atomic fluorescence signal under a
particular condition from all sites of the array (with optical pumping performed
beforehand for $m_F = -3/2$ imaging), we then generate \updated{$200$} bootstrap
data sets, each with \updated{$500$} realizations, by sampling from the original
set with replacement. The above calculation is then carried out for each
bootstrap set, and the uncertainty in $\mathcal{F}$ is obtained from the
standard deviation over the bootstrap sets. We find $\mathcal{F} =
\fidelityarray$ with $F_0 = \fidelityDarray$ and $F_1 = \fidelityBarray$ for the
$m_F = -3/2$ imaging condition, and $\mathcal{F} = 0.995(3)$ with $F_0 =
0.997(2)$ and $F_1 = 0.995(4)$ for the $m_F = -1/2$ condition. We also
approximate the tweezer loading probability $p = \fillarray$ this way as the
fraction of bright shots,
\begin{equation}
    \label{fillfrac}
    p
        = \Pr(B)
.\end{equation}
However, because optical pumping is a process separate from loading the tweezer,
we note that $P(B)$ is more accurately $p \times \eta_\t{OP}$; we decouple these
parameters using additional multi-readout sequences that do not perform optical
pumping directly after loading as described below.

The imaging method used for all of the experiments discussed here  is
state-selective, such that photons are scattered from an atom only if the atom
is in one of our selected qubit states. Therefore, the mixture weights $P_0$ and
$P_1$ are generally combinations of probabilities in a space of events described
by two binary degrees of freedom: (a) the internal qubit state of the atom, and
(b) whether an atom whose state is in ${}^1\t{S}_0$--${}^3\t{P}_1$ manifold is
present in the tweezer. Here, (a) is the only relevant degree of freedom and (b)
represents an error channel to which experiments are coupled via atom loss,
either through off-resonant scatter to a dark state outside the
${}^1\t{S}_0$--${}^3\t{P}_1$ manifold or through heating. Due to this, it is
then impossible to determine whether a single given image classified as dark has
an atom in the non-fluorescent qubit state or no atom at all.

To resolve this, we append a final, state-independent measurement (called ``atom
readout'' in the main text) at the end of each experimental sequence, such that
it becomes possible to determine which shots of the sequence contain atoms. Then
filtering out all sequence shots with dark final measurements, $P_0$ and $P_1$
are directly converted to state probabilities with atom loss errors entirely
decoupled from all measurements, and bright-dark classifications on the
single-image level are mapped to qubit state measurements, with bright
corresponding to $\ket{0}$ and dark to $\ket{1}$. For sufficiently high
$\bar{\eta}_\t{surv}$, qubit states can be measured many times in a single shot
of an experiment using this method, at the cost of immediately rejecting a
portion of data recorded in an experiment that grows roughly as $1 -
\eta_\t{surv}^M$ where $M$ is the number of measurements performed. Figure
\ref{double-histogram-raw} shows the results of the first two readouts of Fig.
\ref{Figure3}(c), leaving out post-selection on the final readout.

\begin{figure}[t!]
    \includegraphics[width=\linewidth]{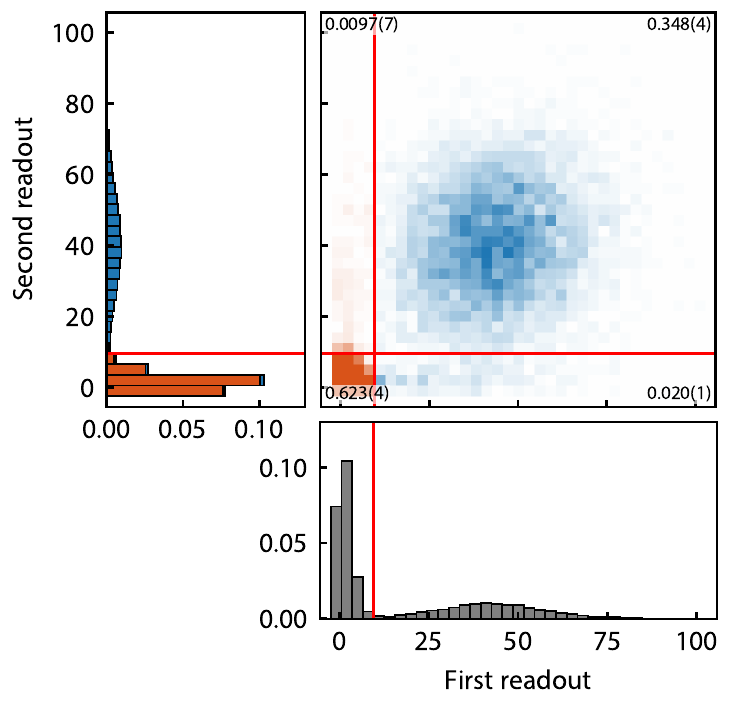}
    \caption{
        \textbf{Depolarization probability study without post-selection on atom
        readout.}
        Photon counts from the first two readouts in the circuit shown in Fig.
        \ref{Figure3}(c) used to study depolarization probability without
        post-selection on the final atom readout. Here, the population in the
        $\t{B}\,\rightarrow\,\t{D}$ depolarization channel is artificially
        inflated by atom loss between these two readouts, as is the population
        in the $\t{D}\,\rightarrow\,\t{D}$ quadrant due to failed loading of the
        tweezer.
        \label{double-histogram-raw}
    }
\end{figure}

Having obtained the discrimination fidelities $F_0$ and $F_1$ above from a
single-measurement sequence assuming a Gaussian mixture model for the
histograms, we now experimentally verify $F_0$ and $F_1$ using an independent
three-measurement sequence inspired by Ref.~\cite{Neumann2010} which we 
describe below. While the state readout fidelity $\mathcal{F}_\t{read}$ and
state initialization fidelity $\mathcal{F}_\t{init}$ are generally coupled, it
is possible to effectively decouple them by using two consecutive measurements.
Explicitly, given a bimodal measurement histogram belonging to the bright
($\ket{0}$) and dark ($\ket{1}$ or empty tweezer) states as shown in Fig.
\ref{Figure1}(d), setting a stricter state-assignment threshold $\theta$ for
$\ket{0}$ ($\ket{1}$) effectively allows one to initialize the state with
perceived increasing probability. Then the resulting histogram of the second
measurement, post-selected on successful initialization via the first
measurement, yields the limiting discrimination fidelities
$\mathcal{F}_\t{read}^\t{B}$ for a tweezer holding an atom in $\ket{0}$
and $\mathcal{F}_\t{read}^\t{D}$ for either an empty tweezer or one holding an
atom in $\ket{1}$.

In practice, this translates to examining the results of a experiment
similar to that shown in Fig.~\ref{Figure3}(c), where the final atom
readout is moved to before the two state readouts. This is done to ensure via
post-selection that we consider only cases where an atom is initially present in
the tweezer, but still include effects arising from imperfect atom survival
between readouts. The method used to analyze results from this sequence
additionally include modified post-selection conditions on the first of the
state readouts.  Holding the threshold $\theta_a$ used for bright/dark
classifications in the first ($a$) measurement fixed at the optimal value
identified by Eq.~\ref{fidelity-def} (i.e. fixing the conditions for detecting
the presence/absence of atoms), we allow the thresholds $\theta_b$ and
$\theta_c$ for the second ($b$) and third ($c$) measurements to vary. Still
post-selecting on a bright initial atom readout, we then examine the results of
the third measurement condition on those of the second as a function of
$\theta_b$ and $\theta_c$ and identify the probabilities $\Pr(B_c | B_a \land
B_b)$ and $\Pr(D_c | B_a \land D_b)$ with fidelities
$\mathcal{F}_\t{read}^\t{B}$ and $\mathcal{F}_\t{read}^\t{D}$, respectively. 

A plot of these quantities as well as their average with $\theta_c$ fixed to its
optimum value is shown in Fig.~\ref{readout-fidelity}. As $\theta_b$ is
increased (decreased), the perceived probability in the $\theta_b \rightarrow
\infty$ ($\theta_b \rightarrow 0$) limit of successfully initializing the state
in $\ket{0}$ ($\ket{1}$) increases, and the second measurement when
post-selected on the success of the first gives the desired state detection
fidelity. We find that $\mathcal{F}_\t{read}^\t{B} = 0.99(1)$ and
$\mathcal{F}_\t{read}^\t{D} = 0.97(2)$, in good agreement with the
discrimination fidelities $F_0$ and $F_1$ measured in
Appendix~\ref{base-fidelity}. We note that this method of estimating
discrimination fidelity is limited by spin-flip events occurring between
readouts. The method additionally relies on full resolution of the limiting
behaviors described above, which requires one to sample events that occur with
vanishing probability with significantly larger data sets. For these reasons, we
choose to use the method based on fitting to Eq.~\ref{mixture-model} and account
for depolarization events in the SPAM models described in Appendix~\ref{spam}.

\begin{figure}[t]
    \includegraphics[width=\linewidth]{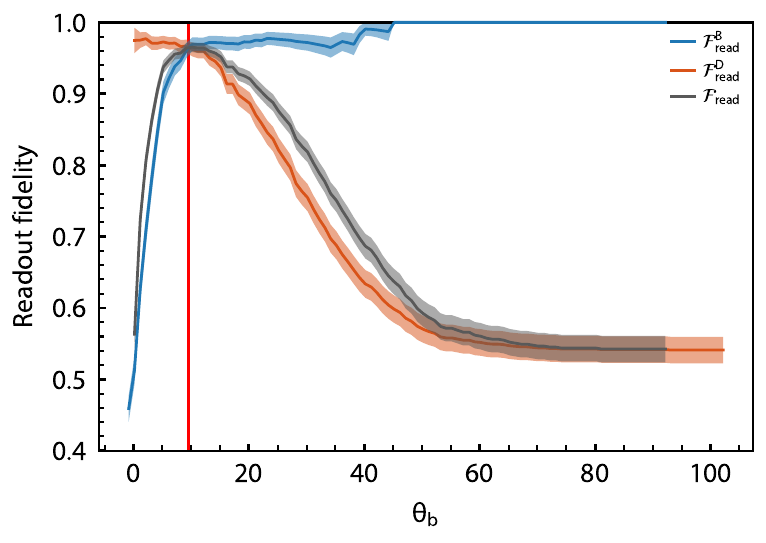}
    \caption{
        \textbf{Validation of discrimination fidelity.}
        Using three consecutive measurement histograms $a, b, c$, with $a$ being
        an atom readout, post-selecting data on measurement $a$ being bright, we
        estimate the readout fidelity by examining the behavior of the
        conditional probabilities $\mathcal{F}_\t{read}^\t{B} = \Pr(B_c | B_a
        \land B_b)$ and $\mathcal{F}_\t{read}^1 = \Pr(D_c | B_a \land D_b)$ as
        well as the average readout fidelity $\mathcal{F}_\t{read} =
        \mathcal{F}_\t{read}^\t{D} \Pr(D_b) + \mathcal{F}_\t{read}^\t{B}
        \t{Pr}(B_b)$ by sweeping the thresholds $\theta_b$ and $\theta_c$ used
        in measurements $b$ and $c$, respectively, while holding that for
        measurement $a$ fixed at the vertical red line, set by the optimization
        of Eq.~\ref{fidelity-def}. We show results for $\theta_c$ fixed to its
        optimum value with the fidelities as a function of $\theta_b$ only. For
        large $\theta_b$, $\ket{0}$ is prepared by $b$ with a perceived
        near-unity probability and subsequently measured in $c$ with fidelity
        \updated{$\mathcal{F}_\t{read}^\t{B} = 0.99(1)$}. For small $\theta_b$,
        $\ket{1}$ is perceived to be prepared with near-unity probability
        instead and measured with fidelity \updated{$\mathcal{F}_\t{read}^\t{D}
        = 0.97(2)$}. We note that $\mathcal{F}_\t{read}^\t{D}$ may not have
        fully saturated to its maximal value due to finite sampling of events
        occurring with vanishing probability.
        \label{readout-fidelity}
    }
\end{figure}

\subsection{Atom survival probability}
\label{survival}
The state-averaged probability $\bar{\eta}_\t{surv}$ that an atom remains in its
tweezer after readout is estimated in two parts.  The $\ket{0}$ bright state
survival $\eta_\t{surv}^\t{B}$  is measured by performing two atom readouts $a$
and $b$ -- of either sort shown in Fig.~\ref{Figure2} -- in quick succession,
post-selecting on a bright $a$ readout, and taking the fraction of those with
bright $b$ readout as well,
\begin{equation}
    \label{survB}
    \eta_\t{surv}^\t{B}
        = \Pr(B_b | B_a)
        = \frac{\Pr(B_a \land B_b)}{\Pr(B_a)}
.\end{equation}
Uncertainty in this measurement is estimated using a similar bootstrapping
procedure to the one for discrimination fidelity described above, wherein $2500
\times 5$ total realizations of this two-readout sequence are again used to
generate $200$ bootstrap data sets, each with $500$ realizations,
$\eta_\t{surv}^\t{B}$ calculated for each one, and the standard deviation across
the bootstrap sets taken.

We find that, for the $m_F = -3/2$ ($m_F = -1/2$) imaging condition taken for a
single array site, $\eta_\t{surv}^\t{B} = \survBsingle$ ($\eta_\t{surv}^\t{B} =
0.96(1)$). Note that the probe time for the $-1/2$ case is $20\,\t{ms}$
(versus $12\,\t{ms}$ for the $-3/2$ case), which is the primary explanation for
the lower survival.

As described in the main text, the survival probability
$\eta_\t{surv}^\t{D} = \survDarray$ of the $\ket{1}$ dark state, which cannot be
directly estimated from the above sequence, is estimated based on the measured
lifetime of the $\ket{1}$ state under probe conditions. We measure this using a
more complex experimental sequence comprising an initial atom readout ($a$) and
a $\pi/2$-pulse, followed by a state readout ($b$), a hold under probe
conditions, and a final atom readout ($c$). We extract the lifetime
$\tausurvDarray$ by fitting the probability $P(B_c | B_a \land D_b)$ as a
function of hold time to a decaying exponential. We also use this sequence to
verify the measured value of $\eta_\t{surv}^\t{B}$ via its lifetime using the
same procedure with $P(B_c | B_a \land B_b)$. This yields a bright state
lifetime of $\tausurvBarray$, which is in good agreement with
$\eta_\t{surv}^\t{B}$ above.

\subsection{Optical pumping efficiency}
Optical pumping is performed using a single beam directed onto the atoms along
an axis perpendicular to that of the tweezers in the horizontal plane. The
polarization of the pump beam is set to be linear, with polarization vector also
lying in the horizontal plane, perpendicular to the quantization axis set by a
$50\,\t{G}$ magnetic field and the polarization of the tweezer light. The
frequency of the beam is set to be near-resonant with the ${}^1\t{S}_0
\leftrightarrow {}^3\t{P}_1$ $F = 3/2$, $m_F = -1/2$ transition, such that the
${}^1\t{S}_0$ $\ket{m_F = +1/2} \equiv \ket{1}$ state is pumped to ${}^1\t{S}_0$
$\ket{m_F = -1/2} \equiv \ket{0}$. The pump beam has $1/e^2$ radius $\approx
2\,\t{mm}$ and intensity $\approx 1.3\,I_\t{sat}$, and is applied for
$80\,\t{$\mu$s}$. We measure the optical pumping efficiency $\eta_\t{OP}$ using
the same two-readout sequence as for bright-state atom survival, now
post-selecting on the $b$ readout:
\begin{equation}
    \label{opeff}
    \eta_\t{OP}
        = \Pr(B_a | B_b)
        = \frac{\Pr(B_a \land B_b)}{\Pr(B_b)}
.\end{equation}
From the same bootstrapping procedure, we find $\eta_\t{OP} = \pumpeffsingle$
(single-site best; $\eta_\t{OP} = \pumpeffarray$ array-averaged).

\subsection{Qubit depolarization probabilities}
As stated in the main text and shown in Fig.~\ref{Figure3}(c), we use two
qubit readouts $a$ and $b$ followed by a single atom readout $c$ to study the
qubit depolarization probability. Post-selecting on a bright $c$ readout to
decouple from atom loss errors, we can directly measure the probabilities
$P_\t{depol}^\t{D\,$\rightarrow$\,B}$ and $P_\t{depol}^\t{B\,$\rightarrow$\,D}$
that the qubit state flips from bright ($\ket{0}$) to dark ($\ket{1}$) or
vice-versa between $a$ and $b$:
\begin{align}
    \label{depolDB}
    P_\t{depol}^\t{D\,$\rightarrow$\,B}
        &= \Pr(B_b | D_a \land B_c)
        = \frac{\Pr(D_a \land B_b \land B_c)}{\Pr(D_a \land B_c)}
    \\
    \label{depolBD}
    P_\t{depol}^\t{B\,$\rightarrow$\,D}
        &= \Pr(D_b | B_a \land B_c)
        = \frac{\Pr(B_a \land D_b \land B_c)}{\Pr(B_a \land B_c)}
.\end{align}
We measure $P_\t{depol}^\t{D\,$\rightarrow$\,B} = \depolDBsingle$ and
$P_\t{depol}^\t{B\,$\rightarrow$\,D} = \depolBDsingle$ (single site best;
$P_\t{depol}^\t{D\,$\rightarrow$\,B} = \depolDBarray$ and
$P_\t{depol}^\t{B\,$\rightarrow$\,D} = \depolBDarray$ array-averaged).

\subsection{$\pi$-pulse fidelity}
Finally we estimate the unitary $\pi$-pulse fidelity $F_\pi$ given by the overlap of the ideal $\pi$-pulse $U_{\pi,\text{ideal}}\!=\!-i\sigma_x$ and the real $\pi$-pulse applied $R(\Omega_0,\delta)\!=\!\exp(-i L/2 (\Omega_0 \sigma_x + \delta \sigma_z))$:
\begin{equation}
    \label{pipulsefidelity}
    \begin{aligned}
        F_\pi
            &= \frac{1}{2}\left|\text{tr}
                \left[U_{\pi,\text{ideal}}^\dag R(\Omega_0,\delta)\right]
            \right|
            \\
            &= \left|
                \frac{\Omega_0}{\Omega} \sin\left(\frac{\theta}{2}\right)
            \right| %|\left (\frac{\Omega_0}{\Omega} \right )\sin(\theta/2)|,
    ,\end{aligned}
\end{equation}
where $\theta=\Omega L$ is given by the generalized Rabi frequency
$\Omega=\sqrt{\Omega_0^2+\delta^2}$ and pulse length $L$.

Noting that $F_\pi$ is given directly by the state transition probability
$\eta_\pi=|\langle 0|R(\Omega_0,\delta)|1\rangle|^2=F_\pi^2$, we use the
three-readout sequence shown in Fig.~\ref{spam-graph} to first extract
$\eta_\pi$. As shown in the model (Fig.~\ref{spam-graph}) we define the
state-averaged $\eta_\pi$:
\begin{equation}
    \label{pipulse}
    \begin{aligned}
        \eta_\pi
            &= \frac{1}{2} \Big[
                \Pr(B_b | D_a \land B_c) + \Pr(D_b | B_a \land B_c)
            \Big]
    ,\end{aligned} 
\end{equation}
and similarly define the (state-averaged) $F_\pi$:
% (\sqrt{\eta_\pi^{1\rightarrow0}}+\sqrt{\eta_\pi^{0\rightarrow1}}/)2
\begin{equation}
    \label{pipulsefidelity}
    \begin{aligned}
        F_\pi
            &= \frac{1}{2} \Big[
                \sqrt{\Pr(B_b | D_a\land B_c)} + \sqrt{\Pr(D_b | B_a \land B_c)}
            \Big]
    .\end{aligned} 
\end{equation}
Thus we measure $F_\pi = \pifidelity$, and $\pifidelitycorr$ with SPAM
correction.

\section{State preparation and measurement correction}\label{spam}

\begin{figure*}[t!]
    \centering
    \includegraphics[width=\textwidth]{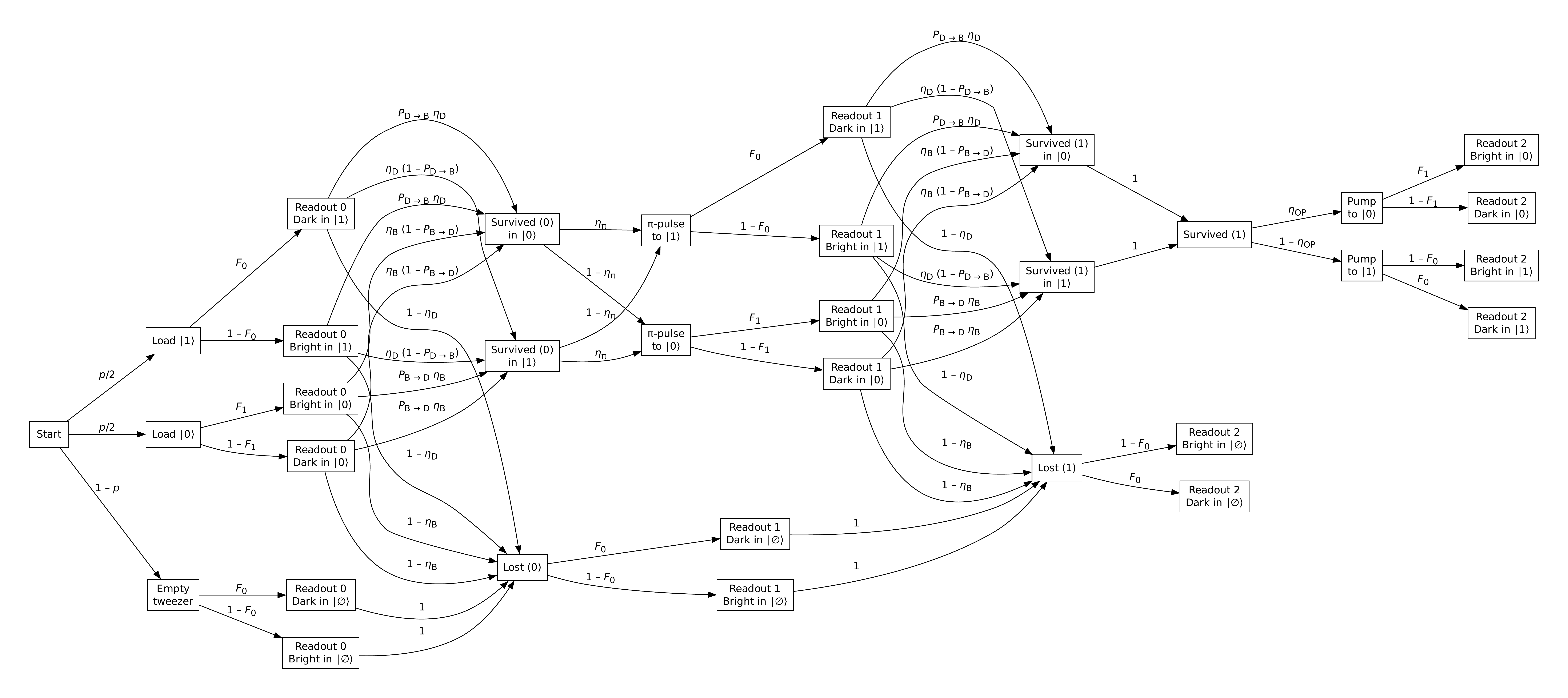}
    \caption{
        \textbf{Example probability graph for SPAM correction.}
        The probability graph for the three-image sequence from
        Fig.~\ref{Figure4}(a) is shown, wherein each node represents a state of
        the system and each edge gives the probability of transitioning to
        another. Every possible combination of readout results is represented as
        the complete path from the ``Start'' node on the left to any of the
        terminal nodes with no outgoing edges on the right, and the total
        probability for any such path is found as the product of all the edge
        weights along it. Here, $\ket{\varnothing}$ is used as a shorthand for
        ``empty tweezer or lost atom,'' and we assume that the qubit state has
        vanishing probability of flipping during readout. Thus, a model for an
        experimentally measurable probability is found by summing the
        probabilities of all paths that satisfy a corresponding set of data
        analysis conditions. For example, the model for the measured $\pi$-pulse
        probability $\spamstar{\eta_\pi}$ is found following Eq.~\ref{pipulse}
        as the sum over probabilities of all paths passing through a ``Bright''
        Readout 0 node and a ``Dark'' Readout 1 node (or vice-versa) as well as
        a ``Bright'' Readout 2 node, divided by that for all paths terminating
        in a ``Bright'' Readout 2 node. This model is set equal to the
        experimentally measured value of $\spamstar{\eta_\pi}$ and used for the
        simultaneous correction of all relevant parameters as described in
        Appendix~\ref{spam}. Generation of models from other graphs is done
        similarly, following Eqs.~\ref{fillfrac}--\ref{depolBD}.
        \label{spam-graph}
    }
\end{figure*}

In experimental sequences, we infer the presence or internal state of atoms by
mapping photon counts onto a binary space and thereby extracting desired
quantities like readout survival probability, qubit depolarization probability,
and $\pi$-pulse fidelity. However, imperfections in tweezer loading probability,
discrimination fidelity, optical pumping efficiency, and indeed qubit
depolarization can produce errors in these measurements. State preparation and
measurement (SPAM) correction attempts to isolate quantities of interest from
sources of error based on a constructed model of possible subprocesses that may
occur in each experimental sequence and estimated probabilities for each of
those subprocesses. Generically, we model a quantity $\spamstar{X}$ measured
directly from experimental data as a function of the isolated ``true'' value of
the quantity $X$ as well as some number of other relevant quantities. A
set of models and experimentally measured values can be used to completely
constrain all other such models, and the system can then be inverted to obtain
SPAM-corrected values for all quantities simultaneously.

Directed, weighted, acyclic graphs provide a convenient language to
describe the various states of the experimental system as it progresses through
a given sequence as well as the measurement outcomes it can produce, and allow
for easy generation of SPAM correction models through well-known graph traversal
algorithms. In these graphs, each node represents a state of the system and each
weighted edge gives the probability of transitioning to another. Every
experimental trajectory is represented as the complete path from a given start
node to any terminal node with no outgoing edges, and the total probability that
the trajectory is realized is the product of the edge weights in the path. An
example graph is shown for the sequence we use to measure the $\pi$-pulse
probability $\eta_\pi$ in Fig.~\ref{spam-graph}, and in general models are
generated by summing over the probabilities of trajectories that follow
Eqs.~\ref{fillfrac}--\ref{pipulse}.

We use a number of these graphs, one for each experimental sequence
described in Appendix~\ref{analysis}, to produce models for all relevant
quantities -- tweezer loading probability $p$, depolarization probabilities
$P_\t{depol}^\t{D\,$\rightarrow$\,B}$ and $P_\t{depol}^\t{B\,$\rightarrow$\,D}$,
survival probabilities $\eta_\t{surv}^\t{B}$ and $\eta_\t{surv}^\t{D}$, optical
pumping efficiency $\eta_\t{OP}$, and $\pi$-pulse fidelity -- and solve the
experimentally constrained system numerically. Uncertainties are estimated by
setting each measured value to its own uncertainty bounds independently,
re-solving the system under all $2^n$ such combinations, and taking the RMS
deviation of this solution set from the nominal corrected values. We note a
corrected fill fraction of $p = \fillarraycorr$ (array-averaged) with the rest
of the SPAM-corrected values are listed in the main text or in
Table~\ref{numbers}.

We note that, here, the state discrimination fidelities $\mathcal{F}$, $F_0$,
and $F_1$ are deliberately left uncorrected in order to simplify calculations:
Strictly, the treatment of the depolarization probabilities here is as the
probability of a bit-flip occurring after a readout is performed, which neglects
the possibility that the flip occurs during the readout itself. We note in this
scenario, however, that because the atom would not scatter photons for the
complete duration of the imaging pulse, the number of photons scattered during
readout would deviate significantly from the means of either the $N = 0$ or $N =
1$ components identified in Eq.~\ref{mixture-model} and be detectable as a count
somewhere in overall count distribution between the two peaks. Since such an
event would directly affect our measured discrimination fidelities, SPAM
correction using uncorrected fidelities therefore constitutes correction for
both discrimination infidelity and any qubit flip errors that occur during a
single readout.

\section{Estimation of $T_1$ and $T_2^*$}\label{T1T2*}
Measurement of the array-averaged $T_1$ depolarization lifetimes is often
hampered by atom loss. When using e.g. a state-dependent blow-away pulse to
measure qubit state populations~\cite{Jenkins2022}, atom loss directly confounds
the measurements from which state populations are inferred. Our ability to
de-couple atom loss from state measurement allows us to measure these quantities
in a straightforward manner. In Figs.~\ref{lifetimes}(a) and \ref{lifetimes}(b),
we show the circuits used to measure $T_1$ and $T_2^*$, respectively. Both
circuits are similar to that shown in Fig.~\ref{Figure3}(c) used to measure
depolarization probability during readout, with two state readouts followed by
an atom readout for post-selection, except for the inclusion of a variable
holding period and $\pi/2$ pulses in the $T_2^*$ case.

\begin{figure}[t]
    \includegraphics[width=0.9\linewidth]{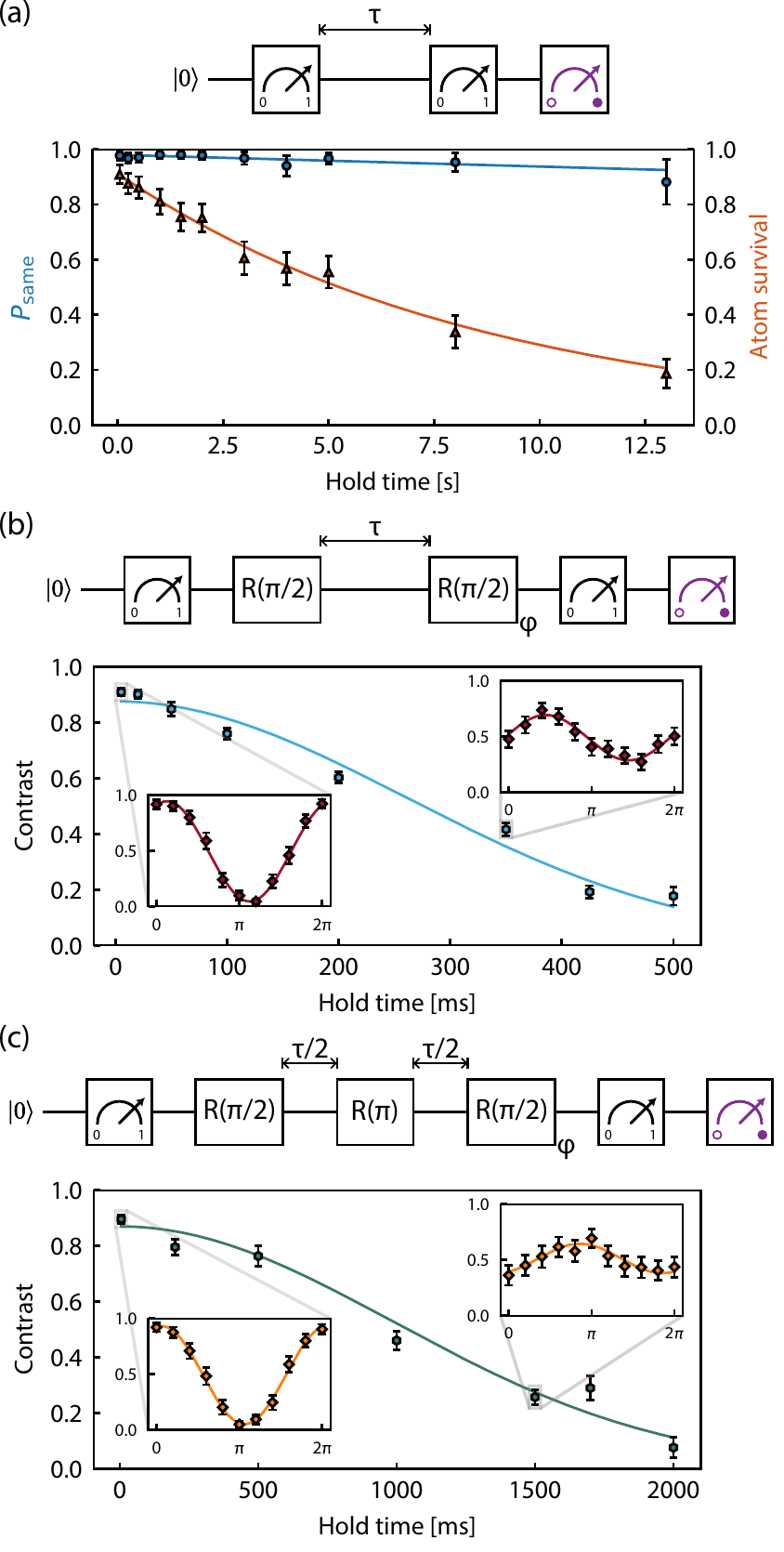}
    \caption{
        \textbf{
            Measurement of array-averaged $T_1$, $T_2^*$,
            and $T_2^\t{echo}$.
        }
        (a) The circuit used to measure dark-state survival and the qubit
        depolarization time, $T_1$. This circuit is identical to that shown in
        Fig.~\ref{Figure3} except for the addition of a holding time $\tau$
        between the first two readouts. The qubit is also initialized in
        $\ket{0}$ here in order to measure both $P_\t{same}$ (blue circles) and
        the atom survival fraction (orange triangles), which are plotted below
        as a function of $\tau$. The atom survival lifetime is measured to be
        $\tausurvvac$. $T_1$ is estimated to be $230(50)\,\t{s}$, but our data
        extends only to $13\,\t{s}$. (b) The circuit used to measure the qubit
        dephasing time, $T_2^*$. Here, the qubit is rotated to the equator for a
        time $\tau$ before a $\pi/2$ pulse with phase $\varphi$ relative to the
        first is applied and the qubit state is measured. The resulting fringes
        in $P_\t{same}$ are plotted as a function of $\varphi$ in the insets,
        and the contrast of the fringes is plotted as a function of the hold
        time $\tau$ in the main plot. The decay of the contrast follows a
        Gaussian profile, the $1/e$ time of which we measure as $T_2^* =
        \taudephase$. (c) The same circuit except with a $\pi$ pulse in
        the middle of the precession time. This spin echo sequence gives
        $T_2^\t{echo}=\taudephaseecho$.
        \label{lifetimes}
    }
\end{figure}

In Fig.~\ref{lifetimes}(a), we initialize in the $\ket{0}$ qubit state via
optical pumping. This effectively turns the first state readout into an atom
readout, which allows us to simultaneously measure atom survival as the number
of shots that are bright in the final atom readout as a fraction of those bright
in the initial readout as well. The atom survival lifetime is obtained by
fitting to a decaying exponential, giving a dark lifetime of $\tausurvvac$.
$T_1$ is similarly measured via the time dependence of the probability
$P_\t{same}$ that the two state readouts give the same result. We estimate that
$T_1=230(50)$ s, but we note that our data only extends to 13 s. We expect that
$\ket{0} \rightarrow \ket{1}$ and $\ket{1} \rightarrow \ket{0}$ depolarization
should have the same rate.

In Fig.~\ref{lifetimes}(b), we measure the array-averaged $T_2^*$ dephasing
time. We again initialize in $\ket{0}$ and apply a $\pi/2$-pulse to rotate the
qubit state to the equator of the Bloch sphere, followed by a variable holding
time and a second $\pi/2$ with variable phase relative to the first. Varying the
relative phase between the pulses leads to fringes in the probability
$P_\t{same}$ that the two state readouts give identical results. $T_2^*$ is
extracted from the time dependence of the contrast of these fringes as the $1/e$
time of a Gaussian profile fit to the data, $T_2^* = \taudephase$. We believe
that this is limited by ambient magnetic field noise in the lab, rather than the
coil servo system. Indeed, we measure the $T_2^*$ at other fields and
find a minimal trend with field: for $\{30, 58, 90\}$ G, we observe
$T_2^*=\{0.39(1), 0.37(1), 0.31(1)\}$ sec. All values range within
$\approx25$\%, yet the naive estimate is that B-field noise should grow
proportionately with the field, for which a range $\approx3$ times larger would
be expected. This data is corroborated by our measurement of
$\approx10\,\t{mG}_\t{pp}$ noise below $1\,\t{kHz}$ with our coils turned off.
We note that noise at higher frequencies has negligible effect on qubit
dynamics, considering our $\pi$-pulse time of $\approx 20\,\t{ms}$.
Additionally, we perform a spin echo sequence to mitigate field noise. As shown
in Fig.~\ref{lifetimes}(c), adding a $\pi$-pulse during the center of the Ramsey
dark time, the coherence is extended by $\approx4\times$ to
$T_2^\t{echo}=\taudephaseecho$.

\section{Real-time feedforward architecture}\label{feedforward-app}
Feedforward to the experiment for qubit $X$ rotations is done by processing
scattered photon counts from images taken by the EMCCD in real time and using
the subsequent bright/dark classification to determine whether the AC magnetic
coils (see Appendix \ref{acfield}) should be driven. More specifically, the AC
current used to drive the coils is generated by a RIGOL DG822 RF source whose
output is gated by an input TTL signal that normally comes directly from the
National Instruments PCIe-7820 board housed in the experiment control computer
(see Appendix \ref{apparatus}). When performing feedforward of the kind
described in Sec. \ref{feedforward}, we insert a switch on the TTL line that is
controlled by an Arduino Uno microcontroller, which is programmed to convert an
ASCII string input over USB from the image-processing computer to a simple
digital HIGH/LOW voltage for the switch. The feedforward logic shown in Fig.
\ref{Figure7} is done by software run by the image-processing computer before
signals are sent to the microcontroller. We note that the time required to
process an image, perform the appropriate logic for the feedforward circuit,
send the signal to the microcontroller, and set the state of the switch
typically adds about $70\,\t{ms}$ per feedforward event to the total
experimental sequence time. This is comparable to our current combined readout
and pulse time, but we expect that this can be significantly reduced through
optimized software and specialized hardware.

\bibliography{library}

\end{document}